\documentclass[11pt]{article}

\pdfoutput=1

\usepackage{amssymb}
\usepackage{amsmath}
\usepackage{subeqnarray}
\usepackage[margin=1.35in]{geometry}
\usepackage{color}

\allowdisplaybreaks[3]

\makeatletter
\@addtoreset{equation}{subsection}
\makeatother

\newcommand{\ams}{${}^1${\it Institute for Theoretical Physics\\
University of Amsterdam, \\
Science Park 904, Postbus 94485, 1090 GL Amsterdam, The Netherlands} 
{\tt M.Smolic@uva.nl}}
\newcommand{\auth}{{\large Milena Smolic${}^1$}}

\def\dalemb#1#2{{\vbox{\hrule height .#2pt
        \hbox{\vrule width.#2pt height#1pt \kern#1pt
                \vrule width.#2pt}
        \hrule height.#2pt}}}

\def\cA{{\cal A}}

\newcommand{\bsea} {\begin{subeqnarray}}
\newcommand{\esea} {\end{subeqnarray}}

\newcommand{\ba}{\beta}

\newcommand{\ud}{\mathrm{d}}

\newcommand{\sig}{\sigma}
\def\0{{\sst{(0)}}}
\def\3{{\sst{(3)}}}
\def\4{{\sst{(4)}}}
\def\5{{\sst{(5)}}}
\def\6{{\sst{(6)}}}
\def\7{{\sst{(7)}}}
\def\8{{\sst{(8)}}}
\def\n{{\sst{(n)}}}

\def\ep{\epsilon}

\def\half{{\textstyle{1\over2}}}

\let\a=\alpha \let\b=\beta  \let\d=\delta \let\e=\epsilon
\let\z=\zeta \let\h=\eta   \let\k=\kappa

\let\m=\mu \let\n=\nu \let\x=\xi \let\r=\rho \let\vp=\varphi
\let\s=\sigma \let\t=\tau  \let\f=\phi  \let\y=\psi
\let\w=\omega

\def\nn{\nonumber} \def\bd{\begin{document}} \def\ed{\end{document}}
\def\ds{\documentstyle} \let\fr=\frac \let\bl=\bigl \let\br=\bigr
\let\Br=\Bigr \let\Bl=\Bigl
\let\bm=\bibitem
\let\na=\nabla
\let\pa=\partial \let\ov=\overline
\newcommand{\be}{\begin{equation}}
\newcommand{\ee}{\end{equation}}
\newcommand{\ch}{\mathrm{\cosh}}
\newcommand{\sh}{\mathrm{\sinh}}

\def\ba{\begin{array}}
\def\ea{\end{array}}
\def\ft#1#2{{\textstyle{{\scriptstyle #1}\over {\scriptstyle #2}}}}
\def\fft#1#2{{#1 \over #2}}
\def\del{\partial}
\def\sst#1{{\scriptscriptstyle #1}}
 \def\oneone{\rlap 1\mkern4mu{\rm l}}
\def\ie{{\it i.e.\ }}
\def\via{{\it via}}
\def\semi{{\ltimes}}
\def\str{{\rm str}}
\def\Dm{{{D_{\sst{max}}}}}
\def\vac{ \left | 0 \right \rangle }
\def\kvac{ \left | k \right \rangle }

\def\sp{\; \; \;}

\def\bol{ \left | B (p^+) \right \rangle}
\def\bo1{ \left | B^0 (p^+) \right \rangle}

\def\bolt{ \left | B (p^+) \right \rangle_{\t}}

\def\boxl{ \left | B (x^-) \right \rangle}

\def\<{ \langle }
\def\>{ \rangle }
\def\({ \left ( }
\def\){ \right ) }

\def\vf{\varphi}

\def\ls{{(l,0)}}
\def\lv{{(l,\pm1)}}
\def\lt{{(l,\pm2)}}

\def\lse#1{{(l_{#1},0)}}
\def\lve#1{{(l_{#1},\pm1)}}
\def\lte#1{{(l_{#1},\pm2)}}

\def\lsg#1{{5(l_{#1},0)}}
\def\lvg#1{{5(l_{#1},\pm1)}}
\def\ltg#1{{5(l_{#1},\pm2)}}

\def\lsi#1{{5{(#1,0)}}}
\def\lvi#1{{5{(#1,\pm1)}}}
\def\lti#1{{5{(#1,\pm2)}}}

\def\lsr#1{{1{(#1,0)}}}
\def\lvr#1{{1{(#1,\pm1)}}}
\def\ltr#1{{1{(#1,\pm2)}}}

\def\cD{{\cal D}}
\def\cE{{\cal E}}
\def\cF{{\cal F}}
\def\cG{{\cal G}}
\def\cH{{\cal H}}
\def\cK{{\cal K}}
\def\cO{{\cal O}}
\def\cP{{\cal P}}
\def\cQ{{\cal Q}}
\def\cR{{\cal R}}
\def\cS{{\cal S}}
\def\cT{{\cal T}}
\def\cU{{\cal U}}
\def\cV{{\cal V}}
\def\cW{{\cal W}}

\newcommand{\nono}{\nonumber}
\newcommand{\dtilde}[1]{\tilde{\tilde{#1}}}
\newcommand{\hatb}[1]{\hat{\ov{#1}}}
\newcommand{\hatt}[1]{\hat{\tilde{#1}}}
\newcommand{\emnr}{{e_\m}^{\n\r}}
\newcommand{\sub}[1]{\phantom{}_{(#1)}\phantom{}}
\newcommand{\rt}{\tilde{\r}}

\newcommand{\comment}[1]{}

\def\hna{\hat{\na}}

\newcommand{\hsp}{\hspace{0.5cm}}

\newcommand{\ho}[1]{$\, ^{#1}$}
\newcommand{\hoch}[1]{$\, ^{#1}$}
\newcommand{\bea}{\begin{eqnarray}}
\newcommand{\eea}{\end{eqnarray}}
\newcommand{\ra}{\rightarrow}
\newcommand{\lra}{\longrightarrow}
\newcommand{\Lra}{\Leftrightarrow}
\newcommand{\ap}{\alpha^\prime}
\newcommand{\bp}{\tilde \beta^\prime}
\newcommand{\tr}{{\rm tr} }
\newcommand{\Tr}{{\rm Tr} }
\newcommand{\NP}{Nucl. Phys. }

\begin{document}

\vspace{15pt}

\begin{center}

{\Large \bf Holography and hydrodynamics for EMD theory with two Maxwell fields}

\vspace{20pt}

\auth

\vspace{15pt}

\vspace{8pt}

{\ams}

\vspace{15pt}

\underline{ABSTRACT}
\end{center}
We use `generalized dimensional reduction' to relate a specific Einstein-Maxwell-Dilaton (EMD) theory, including two gauge fields, three neutral scalars and an axion, to higher-dimensional AdS gravity (with no higher-dimensional Maxwell field). In general, this is a dimensional reduction over compact Einstein spaces in which the dimension of the compact space is continued to non-integral values. Specifically, we perform a non-diagonal Kaluza-Klein (KK) reduction over a torus, involving two KK gauge fields. Our aim is to determine the holographic dictionary and hydrodynamic behaviour of the lower-dimensional theory by performing the generalized dimensional reduction on AdS. We study a specific example of a black brane carrying a wave, whose universal sector is described by gravity coupled to two Maxwell fields, three neutral scalars and an axion, and compute the first order transport coefficients of the dual theory. In these theories $\hat{\z}_s / \hat{\eta} < 2\(1/(d-1)-\hat{c}_s^2\)$, where $\hat{c}_s$ is the speed of sound, violating a conjectured bound, but an alternative bound is satisfied.
\newpage
\tableofcontents
\addtocontents{toc}{\protect\setcounter{tocdepth}{2}}

\section{Introduction}

Einstein-Maxwell-Dilaton (EMD) theories have a wide appeal because their field content is suitable for describing finite charge density systems, possibly exhibiting condensates \emph{via} holography \cite{Taylor:2008tg,Gubser:2009qt,Goldstein:2009cv,Cadoni:2009xm,Chen:2010kn,Charmousis:2010zz,Lee:2010qs,Lee:2010ii,Perlmutter:2010qu,Liu:2010ka,Goldstein:2010aw,Cadoni:2011kv,Iizuka:2011hg,Gouteraux:2011ce,Hartnoll:2011fn}, which has obvious uses within condensed matter theory (CMT) in particular. 

We aim in this paper to provide a holographic dictionary for a class of EMD theories, specifically containing two Maxwell fields, three neutral scalars and an axion. A similar analysis was already done in \cite{ouroldpaper}, in which we listed and discussed the cases where the EMD theory can be oxidized to a higher-dimensional AdS-Maxwell theory, and then specialized to the EMD case involving one Maxwell field and two neutral scalars. The motivations to study an EMD theory with two Maxwell fields and an axion are numerous. Firstly, a family of theories that could be obtained from AdS was discussed in \cite{ouroldpaper}, and it would be interesting to understand whether the generalization to many gauge fields is trivial or if new issues arise. Studying the case of two gauge fields is a first step in this direction. The case of many gauge fields may have applications in future, among them the possible holographic description of imbalanced superconductors. In \cite{ouroldpaper} it was also noted that the bound suggested in \cite{Buchel:2007mf} for the bulk to shear viscosity ratio is violated, and a new modified bound was introduced, which was indeed satisfied by the system at hand. We also use our transport coefficient results in this paper to verify this new modified bound. Furthermore, in the context of AdS/CMT, it may also be possible to make a link to the systems studied in \cite{lisa}, since they have a similar structure to the system dealt with here. Finally, due to the presence of the axion in our theory, our analysis may allow for the holographic modelling of axion physics. This has many applications, among them in cosmology, where axions are considered a possible dark matter candidate.

Setting up holography for such EMD systems is nontrivial in general since a large proportion of solutions are not asymptotically AdS. We bypass the difficulties associated with the standard method for deriving the holographic dictionary (see \cite{Skenderis:2002wp,Papadimitriou:2004ap} for reviews) by turning to a rather neat `trick', namely `generalized dimensional reduction'. This method allows us to start from a theory whose holographic dictionary is known, and infer the holographic dictionary of the theory related to it \emph{via} such a reduction.

This generalized dimensional reduction needs to be consistent, which means that all lower-dimensional theory solutions must also be solutions of the higher dimensional theory. This allows us to infer, among other things, the structure of the lower-dimensional field equations from the higher-dimensional counterpart. This is particularly important since one of the main ingredients in setting up holography is understanding the asymptotic structure of the field equations. The term `generalized' refers to the additional requirement that the reduced theory depends smoothly on the dimension of the compactification manifold, allowing one to continue this parameter to be any real number. In particular, this generalized reduction method was successfully applied to higher-dimensional AdS gravity coupled to matter fields in \cite{Kanitscheider:2009as}, to obtain the holographic setup for lower-dimensional gravity coupled to a scalar field with exponential potential, which itself is associated with the near-horizon limit of non-conformal branes \cite{Itzhaki:1998dd,Boonstra:1998mp}. A diagonal reduction over a $\mathbf T^{2\s-d}$ torus was used. In this paper we want to set up holography for a lower-dimensional theory with two Maxwell fields. We do so by replacing the diagonal torus reduction by a general non-diagonal reduction. In \cite{ouroldpaper} this was done to yield one Maxwell field in the lower-dimensional theory, and we simply extend this by a further reduction to obtain the additional Maxwell field. An intermediate step in the process will give rise to the axion: the Maxwell field introduced by the first reduction is itself reduced to ultimately yield the axion.

A particularly pleasing property of generalized dimensional reduction is that it yields theories, from higher-dimensional AdS gravity (possibly coupled to a Maxwell field), which have known non-extremal black hole solutions (with nontrivial scalar and Maxwell fields). These solutions are not only complicated in structure, but also often have atypical asymptotic behaviour, so studying their holographic setup from scratch is quite an intricate affair. However, we may deduce much about these solutions due to their simple higher-dimensional origins. Among other things, they satisfy all expected thermodynamic identities as does their AdS analogue \cite{Papadimitriou:2005ii}, it becomes relatively straightforward to compute conserved charges \emph{via} the holographic stress tensor and holographic conserved current, and we may also gain an insight into the lower-dimensional hydrodynamic regime. We explicitly use the generalized dimensional reduction to derive the relevant transport coefficients for our lower-dimensional theory.

This paper is organised as follows. In section \ref{sec:EMDtwoMax} we perform a generalized dimensional reduction on AdS gravity to obtain the desired lower-dimensional theory containing three neutral scalars, an axion and two Maxwell fields, and its corresponding holographic dictionary. In section \ref{sec:bbtwoMax}, we study a specific example of a black brane carrying a wave (whose universal sector contains precisely the fields mentioned above), and use the dimensional reduction to obtain a description of its hydrodynamic regime by calculating the relevant transport coefficients. In the appendices we write down the equations of motion of the reduced theory with two gauge fields, give an explicit check of the quantities sourced by the non-normalizable modes of fields within the theory, derive the transport coefficient relations in a two charge hydrodynamic system, provide details regarding the computation of transport coefficients, give a derivation for the relation used as an independent check of our result for the bulk to shear viscosity ratio, and finally provide the explicit calculation of this check.

\section{Holography for EMD theory with two Maxwell fields }\label{sec:EMDtwoMax}

In \cite{ouroldpaper} we started from the higher-dimensional AdS-Maxwell action and wrote down the $(d+1)$-dimensional action obtained \emph{via} a general non-diagonal torus reduction over $\mathbf T^{(2 \sig-d)}$ involving ${\cal M}$ Kaluza-Klein gauge fields. We also worked out explicitly the case corresponding to a lower-dimensional theory with one Maxwell field, starting from higher-dimensional AdS gravity. In this section we will restrict ourselves to an analysis involving another particular example of this general case, where upon reduction of higher-dimensional AdS gravity (with cosmological constant but no higher-dimensional Maxwell field), we obtain lower-dimensional EMD theory with two Maxwell fields, three Kaluza-Klein scalars and an axion. We may perform this reduction in two steps. The first involves a torus reduction of (\ref{AdS_up}) involving one Kaluza-Klein gauge field, as done explicitly in \cite{ouroldpaper}. We then reduce the resulting intermediate action, including a second Kaluza-Klein gauge field. The axion comes from the reduction of the Maxwell field present in the intermediate action due to the first reduction. 

Here, we will provide the generalized Kaluza-Klein reduction map, write down the resulting lower-dimensional action, and then move on to deriving the holographic dictionary for the reduced theory. Note that by setting one of the scalar fields $\x$, the gauge field $A_{M}^{(1)}$ as well as the axion $A^{(0)}$ to zero in this section, we recover the results of \cite{ouroldpaper}.

\subsection{Generalized dimensional reduction} \label{subsec:genred}

We begin with a higher-dimensional action without any Maxwell fields, namely Einstein gravity with negative cosmological constant in $(2\s +1)$-dimensions
\be \label{AdS_up}
S_{(2 \s+1)} = L_{AdS} \int \ud^{2 \s+1} x \sqrt{-g_{(2 \s+1)}}
\left[R + 2 \s (2 \s-1)\right]\,,
\ee
where  $L_{AdS} =  \ell_{(2 \sig+1)}^{2\s-1}/(16 \pi G_{2\s+1})$,  $\ell_{(2 \sig+1)}$ is the
AdS radius and we used an appropriate Weyl rescaling to move $\ell_{(2 \sig+1)}$
as an overall constant in the action. We perform the reduction of such an action, ultimately ending up with a lower-dimensional theory with two Maxwell fields, an axion and three scalars. 

The full reduction ansatz for the theory on the torus $\mathbf T^{(2 \sig-d)}$ is given by
\bea
\label{redans}
\ud s_{(2 \sigma +1)}^2 &=& \ud s^2_{(d+1)}(\r,z) + e^{2 \phi_3(\r,z)} \left(\ud y_{2}  - A_N^{(2)} \ud x^N\right)^2 \nn\\
&+& e^{2 \phi_1(\r,z)}\left(\ud y_{1}  - A_{M^{\prime}}^{(1)} \ud x^{M^{\prime}} \right)^2 \nn\\
&+& e^{\frac{2\phi_2(\r,z)}{(2\s-d-2)}} \ud y^a \ud y^a\,,
\eea
where $a=1,\ldots ,(2\s-d-2)$. The coordinates $(y_{1}, y_{2},y^a)$ are periodically
identified with period $2 \pi R$, $x^M=(\rho,z^i)$ with $M=0,\ldots, d$, and $x^{M^{\prime}}=(x^M,y_{2})$. Furthermore, we include the axion $A^{(0)}$ in the reduction of $A_{M^{\prime}}^{(1)}$:
\be
\label{axionred}
A_{M^{\prime}}^{(1)} = \(A_{M}^{(1)},A^{(0)}\)\,.
\ee
Using a more natural combination of the scalars, namely
\bea\label{psizeta}
&&\vp = \phi_1 + \phi_2\,, \qquad \x = (2 \s - d  -2 ) \f_1 - \f_2\,,\qquad \y=\vp+\f_3\,,\nn\\
&&\qquad \qquad\qquad\qquad \z=(2\s - d -1)\f_3 - \vp\,, \label{phixi}
\eea
the reduced action is calculated to be
\bea
S_{(d+1)} &=&  L
\int \ud^{d+1} x \sqrt{-g_{(d+1)}} e^{\y} \left[ R_{(d+1)} + \frac{2\s-d-1}{(2 \s -d)} (\partial \y)^2 \right. \nn\\
&& \left. - \frac{1}{(2 \s -d)(2\s-d-1)} (\partial \z)^2 - \frac{1}{(2 \s -d-1)(2\s-d-2)} (\partial \x)^2 \right.  \nn\\
&& \left. -\frac{1}{4} e^{\frac{2 (\y + \z)}{(2 \s-d)}} F_{MN}^{(2)} F^{(2)MN} -\frac{1}{4}e^{\frac{2(\x(2\s-d)+(2\s-d-1)\y-\z)}{(2\s-d)(2\s-d-1)}}\times \right. \nn\\
&&\left. \times\(F_{MN}^{(1)}F^{(1)MN} + 4\pa^M A^{(0)}A^{(2)N}F_{MN}^{(1)} -2\(A^{(2)M}\pa_M A^{(0)}\)^{2} \right. \right. \nn\\
&&\qquad\qquad \left. \left. +2\(e^{\frac{-2 (\y + \z)}{(2 \s-d)}}+A_{M}^{(2)}A^{(2)M}\)\(\pa A^{(0)}\)^{2} \)  +2\s(2\s-1)\right]\,, \nn\\
&&\mbox{\ \ } \label{redactfull}
\eea
where $L = L_{AdS} (2 \pi R)^{2\s-d}$, while the reduction ansatz becomes
\bea
\label{redansfull}
\ud s_{(2 \sigma +1)}^2 &=& \ud s^2_{(d+1)}(\r,z) + e^{\frac{2 (\y + \z)}{(2 \s-d)}}  \left(\ud y_{2}  - A_N^{(2)} \ud x^N\right)^2 \nn\\
&+& e^{\frac{2(\x(2\s-d)+(2\s-d-1)\y-\z)}{(2\s-d)(2\s-d-1)}}\left(\ud y_{1}  - A_{M}^{(1)} \ud x^{M} -A^{(0)}\ud y^{(2)}\right)^2 \nn\\
&+& e^{\frac{2\y}{2\s-d}-\frac{2\z}{(2\s-d)(2\s-d-1)}-\frac{2\x}{(2\s-d-1)(2\s-d-2)}} \ud y^a \ud y^a\,.
\eea
We include the equations of motion corresponding to the fields present in (\ref{redactfull}), in appendix \ref{app:eoms}.

We bring the reduced action into Einstein frame by conformally rescaling it using the metric rescaling 
\be\label{metricrescaling}
g_{MN} = e^{-2\y/(d-1)} \bar{g}_{MN}\,,
\ee
and further rescale the scalars as
\bea\label{NormScalars}
&&\y = \sqrt{\frac{(2\s-d)(d-1)}{2(2\s-1)}} \bar{\y}\,,\qquad \z = \sqrt{\frac{(2\s-d)(2\s-d-1)}{2}} \bar{\z}\,, \nn\\ 
&&\qquad\qquad\qquad \mbox{and}\ \ \ \x = \sqrt{\frac{(2\s-d-1)(2\s-d-2)}{2}} \bar{\x}\,, 
\eea
to yield the action in the Einstein frame with canonically normalized scalar kinetic terms,
\bea
S_{(d+1)} &=&  L
\int \ud^{d+1} x \sqrt{-\bar{g}_{(d+1)}} \left[ \bar{R} - \frac{1}{2} (\pa \bar{\y})^2 - \frac{1}{2} (\pa \bar{\z})^2 - \frac{1}{2} (\pa \bar{\x})^2 \right.\nn \\
&& \left. \qquad\qquad - \frac{1}{4} e^{\sqrt{\frac{2(2\s-1)}{(d-1)(2\s-d)}} \bar{\y} + \sqrt{\frac{2(2\s-d-1)}{2\s-d}} \bar{\z} } F_{MN}^{(2)} F^{(2)MN}\right. \nn\\
&& \left.\qquad\qquad- \frac{1}{4}e^{\sqrt{\frac{2(2\s-d-2)}{(2\s-d-1)}}\bar{\x}+\sqrt{\frac{2(2\s-1)}{(d-1)(2\s-d)}}\bar{\y} - \sqrt{\frac{2}{(2\s-d)(2\s-d-1)}}\bar{\z}}\times \right. \nn\\
&& \left. \qquad\qquad\qquad \times \(F_{MN}^{(1)}F^{(1)MN} + 4\pa^M A^{(0)}A^{(2)N}F_{MN}^{(1)} \right. \right.\nn\\
&& \left.\left.\qquad\qquad\qquad\qquad-2\(A^{(2)M}\pa_M A^{(0)}\)^{2} +2A_{M}^{(2)}A^{(2)M}\(\pa A^{(0)}\)^{2} \) \right. \nn\\
&& \left. \qquad\qquad\qquad\qquad-\half e^{\sqrt{\frac{2(2\s-d-2)}{(2\s-d-1)}}\bar{\x}-\sqrt{\frac{2(2\s-d)}{(2\s-d-1)}}\bar{\z}} \(\pa A^{(0)}\)^{2} \right. \nn\\   
&&\left. \qquad\qquad+ 2\s(2\s-1) e^{-\bar{\y} \sqrt{\frac{2(2\s-d)}{(d-1)(2\s - 1)}}} \right]\,.
\label{finact}
\eea
We may only canonically normalize positive kinetic terms, thus the rescaling (\ref{NormScalars}) assumes that $2\s > (d+2)$, since in (\ref{redactfull}) $2\s < (d+2)$ would render the $\x$ kinetic term negative.

\subsection{Holographic dictionary} \label{subsec:hologdic}

In this section we set up the holographic dictionary of our $(d+1)$-dimensional theory (\ref{finact}) using generalized dimensional reduction, as in \cite{ouroldpaper}. The beauty of this approach lies in the fact that the higher-dimensional theory (\ref{AdS_up}) is relatively simple, with the most general asymptotic solution given by
\bea \label{AdS_exp}
\ud s_{(2 \s +1)}^2 &=& \frac{\ud \rho^2}{4 \rho^2} + \frac{1}{\rho} g_{\mu \nu} \ud x^{\mu} \ud x^{\nu}\,, \\
g_{\mu \nu} &=& g_{(0) \mu \nu} + \rho g_{(2) \mu \nu} + \cdots + \rho^{\sigma} \left ( g_{(2 \sigma) \mu \nu} + h_{(2 \sigma) \mu \nu} \log \rho \right ) + \cdots\,,
\eea
where $g_{(0) \mu \nu}$ is the source. The quantities $\Tr g_{(2\s)}$ and $\nabla^{\mu}g_{(2 \sigma) \mu \nu}$ are determined locally in terms of the source and all other coefficients are completely determined, while the logarithmic terms $h_{(2 \sigma)}$ are present if $\sigma$ is an integer. 

Since all solutions of our theory of interest descend from solutions of (\ref{AdS_up}), we simply need to consider the class of asymptotic solutions which has the form of the reduction (\ref{redansfull}) in order to obtain the general asymptotic solution of (\ref{redactfull}).

We begin by writing down the asymptotic expansions of the $(d+1)$-dimensional fields. We then compute the local boundary counterterm action, needed to ensure that the on-shell action is finite. We do this \emph{via} generalized dimensional reduction, which we further use to compute the holographic one-point functions of the lower-dimensional theory. We then use linear combinations of these one-point functions to build the stress-energy tensor, currents and naturally normalized scalar operators of the dual $d$-dimensional field theory. Finally we reduce the higher-dimensional conformal dilatation Ward identity and stress energy tensor conservation equation.

\subsubsection{Asymptotic expansion}

We expand the $(d+1)$-dimensional metric in the usual Fefferman-Graham form, as in (\ref{AdS_exp}),
\bea
\ud s_{(d +1)}^2 &=& \frac{\ud \rho^2}{4 \rho^2} + \frac{1}{\rho} \tilde{g}_{ij}(z,\r) \ud z^{i} \ud z^{j} \nn\\
&=& \frac{\ud \rho^2}{4 \rho^2} + \frac{1}{\rho} \(\tilde{g}_{(0)ij}+\cdots +\r^{\s}\tilde{g}_{(2\s)ij}\)\ud z^{i} \ud z^{j} \,, \label{AdS_exp_metric}
\eea
whilst the scalar fields can be expanded as
\bea \label{AdS_exp_scalars}
e^{\frac{2 \y}{(2 \sigma -d)}} &=& \frac{1}{\rho} e^{\frac{2 \kappa}{(2 \sigma -d)}}\,, \qquad \kappa = \kappa_{(0)} + \rho \kappa_{(2)}
+ \cdots + \rho^{\s} \kappa_{(2\s)}\,, \nn \\
\zeta &=& \zeta_{(0)} + \rho \zeta_{(2)} + \cdots
+ \rho^{\sigma} \x_{(2 \s)}\,, \nn\\
\x &=& \x_{(0)} + \rho \x_{(2)} + \cdots
+ \rho^{\sigma} \x_{(2 \s)}\,,
\eea
the gauge fields as
\bea \label{AdS_exp_gauge}
{ A_i ^{(1)}(\r,z) = A_{i(0)}^{(1)}(z) + \r A_{i(2)}^{(1)}(z) + \cdots + \r^\s A_{i(2\s)}^{(1)}(z) + \cdots }\,, \nn\\
{ A_i ^{(2)}(\r,z) = A_{i(0)}^{(2)}(z) + \r A_{i(2)}^{(2)}(z) + \cdots + \r^\s A_{i(2\s)}^{(2)}(z) + \cdots }\,, 
\eea
and the axion as
\bea\label{AdS_exp_axion}
{ A^{(0)}(\r,z) = A_{(0)}^{(0)}(z) + \r A_{(2)}^{(0)}(z) + \cdots + \r^\s A_{(2\s)}^{(0)}(z) + \cdots }\,.
\eea
We are interested in cases with non-integral $\s$, so our expansions of interest do not contain the logarithmic terms associated with integer $\s$.

\subsubsection{Counterterms and holographic one-point functions}

With the asymptotic solution in hand, we would now like to compute the local boundary counterterms. These are needed to ensure that the on-shell action is finite. Again we turn to generalized dimensional reduction \cite{Kanitscheider:2009as} to simplify the process, and perform the same analysis as in \cite{ouroldpaper}. So, as an example we consider the counterterm action for $1 < \s < 2$, for which we only need two counterterms. Reducing the two most singular counterterms in AdS$_{2\tilde{\s}+1}$ \cite{deHaro:2000xn} to $d$ dimensions, we obtain
\bea
S_{(d)}^{ct} &=&  L
\int_{\rho=\e} \ud^{d} x \sqrt{-\gamma_{d}} e^{\y} \left[ 2(2\s-1) +\frac{1}{2\s-2}\(\hat{R}_{d} - \frac{2\s-d-1}{(2 \s -d)} (\partial \y)^2\right. \right. \nn\\
&& \left. \left. - \frac{1}{(2 \s -d)(2\s-d-1)} (\partial \z)^2 - \frac{1}{(2 \s -d-1)(2\s-d-2)} (\partial \x)^2 \right. \right.  \nn\\
&& \left. \left. -\frac{1}{4} e^{\frac{2 (\y + \z)}{(2 \s-d)}} F_{ij}^{(2)} F^{(2)ij} -\frac{1}{4}e^{\frac{2(\x(2\s-d)+(2\s-d-1)\y-\z)}{(2\s-d)(2\s-d-1)}}\times \right. \right. \nn\\
&&\left. \left. \times\(F_{ij}^{(1)}F^{(1)ij} + 4\pa^i A^{(0)}A^{(2)j}F_{ij}^{(1)} -2\(A^{(2)i}\pa_i A^{(0)}\)^{2} \right.\right. \right. \nn\\
&&\qquad\qquad \left. \left. \left. +2\(e^{\frac{-2 (\y + \z)}{(2 \s-d)}}+A_{i}^{(2)}A^{(2)i}\)\(\pa A^{(0)}\)^{2} \) \)\right]\,. \nn\\
&&\mbox{\ \ } \label{redactfullct}
\eea

Next we would like to compute the holographic one-point functions of our lower-dimensional theory. However, we know the formula for the higher-dimensional one-point function,
\be
\label{conf_dict}
   \<T_{\m\n}\>_{2\s} = \frac{2}{\sqrt{-g_{(0),2\s}}} \frac{ \d S_{ren}}{\d g_{(0)}^{\m\n}} = 2\s L_{AdS}g_{(2\s) \m\n}  + \cdots\,,
\ee
with the actual quantity of interest being
\be\label{littlet}
\<t_{\mu \nu}\>_d \equiv  e^{\k\sub{0}} (2\pi R)^{2\s-d} \<T_{\mu \nu}\>_{2\s}\,, 
\ee
since it takes into account the prefactors $(e^{\k\sub{0}}, (2\pi R)^{2\s-d})$ resulting from the integration over the torus and the change in the metric determinant when going to $d$-dimensions, respectively. The ellipses in (\ref{conf_dict}) again represent terms dependent locally on the source which appear when $\s$ is an integer, since in that instance $g_{(0)\m\n}$ is curved and there is a conformal anomaly.

Thus, as in \cite{ouroldpaper}, we may again simply dimensionally reduce $\<t_{\mu \nu}\>_d$ to obtain the expectation values of the operators in the $d$-dimensional field theory.
The various components are given by
\bea
\<t_{ij}\>_d  &=& 2\s L e^{\k\sub{0}} \left[ \tilde{g}_{(2\s)ij} + 2e^{\frac{2\(\k\sub{0} + \z\sub{0}\)}{2 \s-d}}
\left(A_{(i(0)}^{(2)} A_{j)(2\s)}^{(2)} \right.\right. \nn\\
&&\left.\left. \qquad\qquad\qquad\qquad+ \frac{A_{i(0)}^{(2)}A_{j(0)}^{(2)}}{2\s-d} \left(\k\sub{2\s} + \z\sub{2 \s}\right)\right)  \right. \nn\\
&& \left.+2e^{\frac{2\((2\s-d-1)\k\sub{0}+(2\s-d)\x\sub{0}-\z\sub{0}\)}{(2\s-d)(2\s-d-1)}} \times \right. \nn\\
&& \left.\qquad\qquad\times\( \(\frac{(2\s-d-1)\k\sub{2\s}+(2\s-d)\x\sub{2\s}-\z\sub{2\s}}{(2\s-d)(2\s-d-1)} \)A_{i(0)}^{(1)}A_{j(0)}^{(1)} \right. \right. \nn\\
&& \left. \left. \qquad\qquad\qquad\qquad+ A_{(i(0)}^{(1)} A_{j)(2\s)}^{(1)} \) \right]\,, \nn\\
\< t_{iy_{1}} \>_d   &=& - 2\s Le^{\k\sub{0}} e^{\frac{2\((2\s-d-1)\k\sub{0}+(2\s-d)\x\sub{0}-\z\sub{0}\)}{(2\s-d)(2\s-d-1)}}\times \nn\\
&&\times \left(A_{i(2\s)}^{(1)} +\( \frac{2\((2\s-d-1)\k\sub{0}+(2\s-d)\x\sub{0}-\z\sub{0}\)}{(2\s-d)(2\s-d-1)}\) A_{i(0)}^{(1)}\right)\,,  \label{t_i1_d}\nn\\
&&\mbox{\ }\\
\< t_{iy_{2}} \>_d   &=&  2\s Le^{\k\sub{0}} \left[ -e^{\frac{2\(\k\sub{0} + \z\sub{0}\)}{2 \s-d}}\( \frac{2}{(2\s-d)} \left(\k\sub{2\s} + \z\sub{2 \s}\right)A_{i(0)}^{(2)} +A_{i(2\s)}^{(2)}\) \right. \nn\\
&& \left.\qquad\qquad+e^{\frac{2\((2\s-d-1)\k\sub{0}+(2\s-d)\x\sub{0}-\z\sub{0}\)}{(2\s-d)(2\s-d-1)}} \times \right. \nn\\
&& \left.\qquad\times\( \(\frac{2\((2\s-d-1)\k\sub{2\s}+(2\s-d)\x\sub{2\s}-\z\sub{2\s}\)}{(2\s-d)(2\s-d-1)} \)A_{i(0)}^{(1)}A_{(0)}^{(0)} \right. \right. \nn\\
&& \left. \left. \qquad\qquad\qquad\qquad+ A_{i(2\s)}^{(1)} A_{(0)}^{(0)} + A_{(2\s)}^{(0)} A_{i(0)}^{(1)} \) \right]\,, \nn\\
\< t_{y_{1}y_{1}} \>_d   &=&  4\s Le^{\k\sub{0}} e^{\frac{2\((2\s-d-1)\k\sub{0}+(2\s-d)\x\sub{0}-\z\sub{0}\)}{(2\s-d)(2\s-d-1)}} \times \nn\\
&& \qquad\qquad\qquad\times\(\frac{(2\s-d-1)\k\sub{2\s}+(2\s-d)\x\sub{2\s}-\z\sub{2\s}}{(2\s-d)(2\s-d-1)} \)+\cdots\nn\\
&\equiv& - e^{\frac{2\((2\s-d-1)\k\sub{0}+(2\s-d)\x\sub{0}-\z\sub{0}\)}{(2\s-d)(2\s-d-1)}}\<\cO_3\>_d\,, \nn\\
\< t_{y_{2}y_{2}} \>_d   &=&4\s Le^{\k\sub{0}} \left[ e^{\frac{2\(\k\sub{0} + \z\sub{0}\)}{2 \s-d}}\( \frac{1}{2\s-d}\) \left(\k\sub{2\s} + \z\sub{2 \s}\right) \right. \nn\\
&& \left.\qquad\qquad+e^{\frac{2\((2\s-d-1)\k\sub{0}+(2\s-d)\x\sub{0}-\z\sub{0}\)}{(2\s-d)(2\s-d-1)}} \times \right. \nn\\
&& \left.\qquad\qquad\qquad\times\( \(\frac{(2\s-d-1)\k\sub{2\s}+(2\s-d)\x\sub{2\s}-\z\sub{2\s}}{(2\s-d)(2\s-d-1)} \)A_{(0)}^{(0)\ 2} \right. \right. \nn\\
&& \left. \left. \qquad\qquad\qquad\qquad+ A_{(0)}^{(0)} A_{(2\s)}^{(0)} \) \right] +\cdots\nn\\
&\equiv& -e^{\frac{2\(\k\sub{0} + \z\sub{0}\)}{2 \s-d}} \<\cO_1\>_d \,,\nn\\
\< t_{y_{1}y_{2}} \>_d   &=&-2\s Le^{\k\sub{0}} e^{\frac{2\((2\s-d-1)\k\sub{0}+(2\s-d)\x\sub{0}-\z\sub{0}\)}{(2\s-d)(2\s-d-1)}} \times \nn\\
&& \qquad\qquad\times\( \(\frac{2\((2\s-d-1)\k\sub{2\s}+(2\s-d)\x\sub{2\s}-\z\sub{2\s}\)}{(2\s-d)(2\s-d-1)} \)A_{(0)}^{(0)} \right.  \nn\\
&& \left.  \qquad\qquad\qquad\qquad+A_{(2\s)}^{(0)} \)+\cdots \nn\\
&\equiv& - e^{\frac{2\((2\s-d-1)\k\sub{0}+(2\s-d)\x\sub{0}-\z\sub{0}\)}{(2\s-d)(2\s-d-1)}}\<\cO_4\>_d\,, \nn\\
\< t_{ab} \>_d   &=&  4\s Le^{\k\sub{0}} e^{\frac{2\((2\s-d-1)(2\s-d-2)\k\sub{0}-(2\s-d-2)\z\sub{0}-(2\s-d)\x\sub{0}\)}{(2\s-d)(2\s-d-1)(2\s-d-2)}} \times \nn\\
&& \times\( \frac{1}{(2\s-d)}\k_{(2\s)} - \frac{1}{(2\s-d)(2\s-d-1)}\z_{(2\s)} \right.\nn\\
&&\left. \qquad\qquad- \frac{1}{(2\s-d-1)(2\s-d-2)}\x_{(2\s)}\)\d_{ab}+\cdots\nn\\
&\equiv& -e^{\frac{2\((2\s-d-1)(2\s-d-2)\k\sub{0}-(2\s-d-2)\z\sub{0}-(2\s-d)\x\sub{0}\)}{(2\s-d)(2\s-d-1)(2\s-d-2)}}\<\cO_2\>_d\d_{ab}\,,\label{vev}
\eea
and from these expressions we read off
\bea
\<\cO_1 \>_d &=& - \frac{4\s L}{2\s-d} e^{\k\sub{0}} \left[\k\sub{2\s} + \z\sub{2\s} + e^{\frac{2\(\x\sub{0}-\z\sub{0}\)}{(2\s-d-1)}}\times \right. \nn\\
&&\left.\times \( A_{(0)}^{(0)\ 2} \( \k\sub{2\s} + \frac{(2\s-d)}{(2\s-d-1)}\x\sub{2\s} -\frac{1}{(2\s-d-1)}\z\sub{2\s}  \) \right.\right. \nn\\
&& \left.\left. \qquad+ (2\s-d) A_{(0)}^{(0)}A_{(2\s)}^{(0)} \) \right] +\cdots \,,\nn\\
\<\cO_2 \>_d&=& - \frac{4\s L}{2\s-d} e^{\k\sub{0}}
\left(\k\sub{2\s}-\frac{1}{(2\s-d-1)}\z\sub{2\s} \right.\nn\\
&&\left. \qquad\qquad\qquad- \frac{(2\s-d)}{(2\s-d-1)(2\s-d-2)}\x\sub{2\s}\right) + \cdots\,,  \nn\\
\<\cO_3 \>_d&=& - \frac{4\s L}{2\s-d} e^{\k\sub{0}}
\left(\k\sub{2\s} + \frac{(2\s-d)}{(2\s-d-1)}\x\sub{2\s} -\frac{1}{(2\s-d-1)}\z\sub{2\s}\right) + \cdots\,,  \nn\\
\<\cO_4 \>_d&=& \frac{2\s L}{2\s-d} e^{\k\sub{0}}\( 2A_{(0)}^{(0)}\( \k\sub{2\s} +\frac{(2\s-d)}{(2\s-d-1)}\x\sub{2\s} - \frac{1}{(2\s-d-1)}\z\sub{2\s}\) \right. \nn\\
&& \left. \qquad\qquad\qquad +(2\s-d)A_{(2\s)}^{(0)} \) + \cdots \,. \nn
\eea
In all of the above expressions, the ellipses represent terms containing derivatives of the scalar sources $(\k_{(0)}, \z_{(0)}, \x_{(0)})$ and curvatures of the boundary metric $g_{(0)ij}$.

The reduction has thus yielded seven arbitrarily normalized operators: a symmetric tensor $\<t_{ij}\>_d$, two vectors $\<t_{iy_{1}}\>_d$ and  $\<t_{iy_{2}}\>_d$, as well as four scalar operators. 

\subsubsection{Stress-energy tensor, currents and scalar operators}

With the reduced one-point functions in hand, we now wish to form appropriate linear combinations of them to yield the stress energy tensor $\<\hat{T}_{ij}\>$, currents $\<\hat{J}_{i}^{(1)}\>$ and $\<\hat{J}_{i}^{(2)}\>$, and naturally normalized scalar operators of the dual $d$-dimensional field theory, namely $\<{\cal O}_\psi\>_d$, $\<{\cal O}_\z\>_d$, $\<{\cal O}_\x\>_d$ and $\<{\cal O}_{A^{(0)}}\>_d$. The specific combinations we use become clear when we study the reduction of the higher-dimensional Ward identities in the next section.

We introduce linear combinations of the scalar operators as follows
\bea
\<  \cO_{\y} \>_d &=& \frac{1}{(2\s -d)} \left [ (2 \s - d -2) \<  \cO_{2} \>_d + \<\cO_{1}\>_{d} +2A_{(0)}^{(0)}e^{\frac{2}{(2\s-d-1)}\(\x\sub{0}-\z\sub{0}\)}\<  \cO_{4} \>_d \right. \nn\\
&& \left. \qquad\qquad +\(1+e^{\frac{2}{(2\s-d-1)}\(\x\sub{0}-\z\sub{0}\)}A_{(0)}^{(0)\ 2}  \)\<  \cO_{3} \>_d\right ]\,, \nn\\
\< \cO_{\x} \>_d &=& \frac{1}{(2 \s -d-1)} \left [ \<  \cO_{3} \>_d - \<\cO_{2} \>_{d} \right ], \nn\\
\< \cO_{\z} \>_d &=& \frac{1}{(2 \s -d)} \left [ \<  \cO_{1} \>_d + 2A_{(0)}^{(0)}e^{\frac{2}{(2\s-d-1)}\(\x\sub{0}-\z\sub{0}\)}\<  \cO_{4} \>_d  \right. \nn\\
&&\left. + \frac{1}{(2\s-d-1)} \left[ \( (2\s-d-1)e^{\frac{2}{(2\s-d-1)}\(\x\sub{0}-\z\sub{0}\)}A_{(0)}^{(0)\ 2} -1\)\<  \cO_{3} \>_d \right.\right. \nn\\
&& \left.\left. \qquad\qquad\qquad\qquad -(2\s-d-2)\<\cO_{2} \>_{d} \right] \right], \nn\\
\< \cO_{A^{(0)}} \>_d &=& -  e^{\frac{2}{(2\s-d-1)}\(\x\sub{0}-\z\sub{0}\)} \left [ \<  \cO_{4} \>_d + A_{(0)}^{(0)} \<\cO_{3} \>_{d} \right ]\,,  \label{ops}
\eea
so that
\bea\label{asymform}
\<  \cO_{\y} \>_d &=& -\frac{4\s L}{(2\s-d)}e^{\k_{(0)}}\k_{(2\s)}\,, \nn\\
\< \cO_{\x} \>_d &=& -\frac{4\s L}{(2\s-d-1)(2\s-d-2)}e^{\k_{(0)}}\x_{(2\s)}\,, \nn\\
\< \cO_{\z} \>_d &=& -\frac{4\s L}{(2\s-d)(2\s-d-1)}e^{\k_{(0)}}\z_{(2\s)}\,, \nn\\
\< \cO_{A^{(0)}} \>_d &=& -2\s Le^{\k_{(0)}} e^{\frac{2}{(2\s-d-1)}\(\x\sub{0}-\z\sub{0}\)}A_{(2\s)}^{(0)} \,. 
\eea
Furthermore, we write
\bea
\label{redforms}
&&A_{(0)i}^{(3)}\equiv A_{(0)i}^{(1)} + A_{(0)}^{(0)}A_{(0)i}^{(2)}\,, \nn\\
&&\< \hat{J}_{i}^{(1)} \>_{d} = \< t_{iy_{1}}\>_{d} + A_{(0)i}^{(3)} \<t_{y_{1}y_{1}}\>_{d} + A_{(0)i}^{(2)} \<t_{y_{1}y_{2}}\>_{d} \,, \nn\\
&&\< \hat{J}_{i}^{(2)} \>_{d} = \< t_{iy_{2}}\>_{d} + A_{(0)i}^{(3)} \<t_{y_{1}y_{2}}\>_{d} + A_{(0)i}^{(2)} \<t_{y_{2}y_{2}}\>_{d} \,, \nn\\
&& \< \hat{T}_{ij} \>_{d} = \<t_{ij}\>_{d} + \left(A_{(0) i}^{(3)} \< \hat{J}_{j}^{(1)} \>_d + A_{(0)j}^{(3)} \< \hat{J}_{i}^{(1)} \>_d\right) + \left(A_{(0) i}^{(2)} \< \hat{J}_{j}^{(2)} \>_d + A_{(0)j}^{(2)} \< \hat{J}_{i}^{(2)} \>_d\right) \nn\\
&&\qquad\qquad +A_{(0)i}^{(3)} A_{(0) j}^{(3)} e^{\frac{2\((2\s-d-1)\k\sub{0}+(2\s-d)\x\sub{0}-\z\sub{0}\)}{(2\s-d)(2\s-d-1)}}\<\cO_{3}\>_{d} +A_{(0)i}^{(2)} A_{(0) j}^{(2)}e^{\frac{2\(\k\sub{0} + \z\sub{0}\)}{2 \s-d}} \<\cO_{1}\>_{d}  \nn\\
&& \qquad\qquad+ \( A_{(0)i}^{(3)} A_{(0) j}^{(2)} + A_{(0)i}^{(2)} A_{(0) j}^{(3)}\)e^{\frac{2\((2\s-d-1)\k\sub{0}+(2\s-d)\x\sub{0}-\z\sub{0}\)}{(2\s-d)(2\s-d-1)}}\<\cO_{4}\>_{d} \,,\label{set1}
\eea
so that
\bea
&& \< \hat{J}_{i}^{(1)} \>_{d} = -2 \s L e^{\k\sub{0}} e^{\frac{2\((2\s-d-1)\k\sub{0}+(2\s-d)\x\sub{0}-\z\sub{0}\)}{(2\s-d)(2\s-d-1)}} \( A_{i(2\s)}^{(1)} + A_{(2\s)}^{(0)}A_{i(0)}^{(2)}  \) + \cdots\,, \nn \\
&&\mbox{\ } \label{J1}\\
&& \< \hat{J}_{i}^{(2)} \>_{d} = -2 \s L e^{\k\sub{0}} \( e^{\frac{2\(\k\sub{0} + \z\sub{0}\)}{2 \s-d}} A_{i(2\s)}^{(2)} -e^{\frac{2\((2\s-d-1)\k\sub{0}+(2\s-d)\x\sub{0}-\z\sub{0}\)}{(2\s-d)(2\s-d-1)}} \times \right. \nn\\
&& \left. \qquad\qquad\qquad\qquad\qquad \times \( A_{i(2\s)}^{(1)} + A_{(2\s)}^{(0)}A_{i(0)}^{(2)}  \) A_{(0)}^{(0)} \)+ \cdots\,,  \label{J2}\\
&& \< \hat{T}_{ij} \>_{d} = 2 \s L e^{\k\sub{0}} \tilde{g}_{(2\s)ij} + \cdots\,. \label{T}
\eea
In fact, we may also check that varying the renormalized on-shell action with respect to the appropriate source yields each of these combinations in turn. In appendix \ref{app:inverse_zero_metric} we show explicitly that $\<\hat{J}^{(1)}_i\>_d$ is the current sourced by $A_{(0)}^{(3)i}$, and $\< \hat{J}^{(2)}_i \>_d$ is the current sourced by $A_{(0)}^{(2)i}$. The metric elements of $g_{(0)\mu\nu}$ and its inverse are also supplied in this appendix. 

Now, a distinctive property of the axion is that it enters the action with a derivative, which means that the action remains invariant if one shifts the axion by a constant. Thus, a non-trivial check of the formulas for the one-point functions is that they are all invariant under such a shift. This is true for all of the one-point functions except for $\hat{J}^{(2)}_i$. One can easily see this by looking at the form of (\ref{asymform}) and (\ref{J1}) - (\ref{T}): the only one-point function which has $A_{(0)}^{(0)}$ in it is clearly (\ref{J2}), and under a constant shift in the axion ($A_{(0)}^{(0)} \rightarrow A_{(0)}^{(0)} + c$), it transforms as 
\be
\< \hat{J}^{(2)}_i \>_d \rightarrow \< \hat{J}^{(2)}_i \>_d - c \< \hat{J}^{(1)}_i \>_d\,.
\ee
This is actually consistent with the invariance of the action under the axion shift, and arises due to the transformation of $A_{(0)i}^{(3)}\rightarrow A_{(0)}^{(3)i} + cA_{(0)}^{(2)i}$. More precisely, from the last line in (\ref{genfuncJ}) it follows that the coupling in the action is $A_{(0)}^{(3)i}\hat{J}^{(1)}_{i}+ A_{(0)}^{(2)i}\hat{J}^{(2)}_i$, and under the axion shift this combination remains unchanged.

\subsubsection{Reduced Ward identities}

Beginning with the conformal Ward identity in the $2\s$-dimensional theory,
\be
\<T^\m_\m\>_{2\s}\equiv g_{(0)}^{\mu\nu}\<T_{\mu\nu}\>_{2\s} = \cA_{2\s}\,,
\ee
we use (\ref{metricinverse}) and reduce it to
\bea
&&\<t_{i}^{i}\>_d +2\(A_{(0)}^{(1)i} + A_{(0)}^{(2)i}A_{(0)}^{(0)}\)\<t_{iy_{1}}\>_d + 2A_{(0)}^{(2)i} \<t_{iy_{2}}\>_d \nn\\
&& - \( 1+e^{\frac{2}{(2\s-d-1)}\(\x\sub{0}-\z\sub{0}\)}A_{(0)}^{(0)\ 2}  + e^{\frac{2\((2\s-d-1)\k\sub{0}+(2\s-d)\x\sub{0}-\z\sub{0}\)}{(2\s-d)(2\s-d-1)}} \times \right. \nn\\
&&\left. \qquad \times \(A_{(0)n}^{(1)}A_{(0)}^{(1)n} +A_{(0)n}^{(2)}A_{(0)}^{(2)n}A_{(0)}^{(0)\ 2} +2A_{(0)n}^{(1)}A_{(0)}^{(2)n}A_{(0)}^{(0)}\) \) \<\cO_3\>_d\nn\\
&&-2\( e^{\frac{2}{(2\s-d-1)}\(\x\sub{0}-\z\sub{0}\)}A_{(0)}^{(0)} + e^{\frac{2\((2\s-d-1)\k\sub{0}+(2\s-d)\x\sub{0}-\z\sub{0}\)}{(2\s-d)(2\s-d-1)}} \times \right. \nn\\
&& \left. \qquad\qquad\times \( A_{(0)n}^{(1)}A_{(0)}^{(2)n} + A_{(0)n}^{(2)}A_{(0)}^{(2)n}A_{(0)}^{(0)}\) \) \<\cO_4\>_d\nn\\
&& -\( 1+e^{\frac{2\(\k\sub{0} + \z\sub{0}\)}{2 \s-d}} A_{(0)n}^{(2)}A_{(0)}^{(2)n}\)\<\cO_1\>_d -(2\s-d-2)\<\cO_2\>_d  \nn\\
&& = e^{\k\sub{0}} (2\pi R)^{2\s -d}\cA_{2\s} \equiv \cA_d\,, \nn
\eea
and using (\ref{T}) and (\ref{ops}) we see that this becomes
\be \label{dil_WI}
{\<\hat{T}_{i}^{i}\>_{d} - (2 \s  -d ) \<\cO_{\y}\>_{d} = \cA_{d}}\,.
\ee
Note in particular that the new scalar operators $\cO_{\z}$ and $\cO_{\x}$, and the axion operator $\cO_{A^{(0)}}$ do not contribute to the dilatation Ward identity above.

In addition, reducing the conservation equation for the higher-dimensional stress energy tensor, namely 
\be
\na^\m\<T_{\m\n}\>_{2\s}=0\,,
\ee
yields
 \bea\label{ConsEqT}
&&\tilde{\na}^i \<\hat{T}_{ij} \>_{d} + \pa_j \k\sub{0} \< \cO_\y \>_{d} + \pa_j \z\sub{0} \< \cO_\z \>_{d} + \pa_j \x\sub{0} \< \cO_\x \>_{d}\nn\\
&&+\ \pa_j A_{(0)}^{(0)}\< \cO_{A^{(0)}} \>_{d} + \tilde{F}_{(0)j}^{(3)\ i}\< \hat{J}_{i}^{(1)}\>_{d} + \tilde{F}_{(0)j}^{(2)\ i}\< \hat{J}_{i}^{(2)}\>_{d}= 0\,,
\eea
{and} the divergence equations for two currents
\be \label{ConsEqJ}
{\tilde{\na}^i \< \hat{J}_{i}^{(1)} \>_{d} = 0\,,\ \ \ \mbox{and}\ \ \ \tilde{\na}^i \< \hat{J}_{i}^{(2)} \>_{d} = 0\,. }
\ee
The form of (\ref{ConsEqT}) is a further justification for the combinations (\ref{ops}) and (\ref{set1}) we made in the previous section, since it clearly corresponds to the standard diffeomorphism Ward identity for a theory with stress energy tensor $\hat{T}_{ij}$ in which the other operators are defined in terms of the generating functional $W$ 
\bea
&&\< \cO_{\y} \>_d = - \frac{1}{\sqrt{-g_{(0)}}} \frac{\delta W}{\delta \k\sub{0}}\,,\qquad\< \cO_{\zeta} \>_d = - \frac{1}{\sqrt{-g_{(0)}}} \frac{\delta W}{\delta \zeta\sub{0}}\,, \nn\\
&&\< \cO_{\x} \>_d = - \frac{1}{\sqrt{-g_{(0)}}} \frac{\delta W}{\delta \x\sub{0}}\,, \qquad\< \cO_{A^{(0)}} \>_d = - \frac{1}{\sqrt{-g_{(0)}}} \frac{\delta W}{\delta A_{(0)}^{(0)}}\,,\nn\\
&&\< \hat{J}^{(1)i} \>_d= -\frac{1}{\sqrt{-g_{(0)}}} \frac{\delta W}{\delta A_{(0)i}^{(3)}}\,, \qquad\< \hat{J}^{(2)i} \>_d= -\frac{1}{\sqrt{-g_{(0)}}} \frac{\delta W}{\delta A_{(0)i}^{(2)}}\,. \label{genfuncJ}
\eea
We thus see that the non-normalizable modes of the fields $\(\y,\z,\x,A^{(0)}\)$ indeed source $\(\cO_{\y}, \cO_{\zeta}, \cO_{\x}, \cO_{A^{(0)}}\)$ respectively, whilst $A_{(0)i}^{(3)}$ and $A_{(0)i}^{(2)}$ source the conserved currents $\hat{J}^{(1)i}$ and $\hat{J}^{(2)i}$ respectively. 

\section{Black brane universal sector with two Maxwell fields}\label{sec:bbtwoMax}

We wish to study the universal hydrodynamics of non-conformal branes, and specifically we want to make use of the holographic results obtained in section \ref{sec:EMDtwoMax}, which correspond to action (\ref{redactfull}), so we begin by writing down a black brane solution which carries a wave, and whose universal sector contains the fields we encountered, namely three scalars, an axion and two Maxwell fields. We do this by applying generalized dimensional reduction to a higher-dimensional conformal black brane solution. Now, universal hydrodynamics may be derived by studying the long wavelength fluctuation equations around boosted black brane geometries. Conformal hydrodynamics was derived in this way using the boosted black D3 brane geometry \cite{Bhattacharyya:2008jc}, and it is possible to derive non-conformal hydrodynamics using the boosted black D$p$ brane geometry. Thus, once we have our lower-dimensional boosted brane of interest, we may calculate the transport coefficients corresponding to first-order non-conformal hydrodynamics on the boundary. In \cite{ouroldpaper} we worked out the transport coefficients of a boundary theory dual to a boosted black brane whose universal sector has only one Maxwell field. We may recover all results obtained in \cite{ouroldpaper} by setting $\w_1=0$, $\w_2=\w$, $\tilde{A}_{(0)}^{(1)}=0$ and $\tilde{A}_{(0)}^{(2)}=\tilde{A}_{(0)}$ in the results of this section.

\subsection{Black branes} \label{Gen_br}

In this section we show an explicit realization of the setup discussed in section \ref{sec:EMDtwoMax} for a particular example: a black brane carrying a wave, whose universal sector is described by gravity coupled to two Maxwell fields, three neutral scalars and an axion. Specifically, we show the dimensional reduction involved in producing this system starting from a higher-dimensional conformal black brane solution, and then use the general setup developed in section \ref{subsec:hologdic} to determine the equilibrium thermodynamic quantities of the boundary non-conformal hydrodynamics. In principle, we mirror the analysis of \cite{ouroldpaper}, but additionally perform another Lorentz transformation on the conformal black brane background in order to get it into the desired form.

\subsubsection{Conformal black brane with wave}

As in \cite{ouroldpaper}, our starting point is a (conformal) black brane solution in $(2 \s +1)$ dimensions:
\bea
\label{AdSbbrane}
\ud s^2_{(2\s+1)} &=& \frac{\ud \r^2}{4\r^2 f(\rho)} +\frac1\r\left[-f(\r)\ud t'^2 + \ud y'^2 + \ud y''^2 + \ud z_r \ud z^r+ \ud y_a \ud y^a\right], \nn\\
f(\r) &=& 1 - m^{2\s} \r^\s\,, \nn\\
&&\mbox{\ \ }
\eea
where $(y',y'',y^a, z^r)$ run over all transverse coordinates ($a=d+1, \ldots, 2\s-2$). Note that we highlight the transverse coordinates $y'$ and $y''$ because we will boost along each of them by performing successive Lorentz transformations. 

Now, the metric (\ref{AdSbbrane}) is Einstein with negative curvature when $2 \s $ is an integer, and has an event horizon at $\r=m^{-2}$.  The Hawking temperature $T$ and Bekenstein-Hawking entropy density $s$ are given by
\be \label{entropyAdSbbrane}
T = \frac{ m \sigma}{2 \pi}\,, \qquad s = 4 \pi L_{AdS}  m^{2 \sigma-1}\,.
\ee
We introduce a wave into this metric by performing a first Lorentz transformation $t=\cosh\w_1~t'-\sinh\w_1~y'\,,\,y_1=\cosh\w_1 ~y'-\sinh\w_1~t'$, followed by a second $t''=\cosh\w_2~t-\sinh\w_2~y''\,,\,y_2=\cosh\w_2 ~y''-\sinh\w_2~t$, which gives: 
\bea
\label{bbraneconfdt2}
\ud s^2_{(2\s+1)} &=& \frac{\ud \r^2}{4\r^2 f(\rho)} -  \rho^{-1} K_1(\r)^{-1}K_2(\r)^{-1} f(\r)\ud t''^2 +\ \rho^{-1} \ud z_r \ud z^r+\ \rho^{-1} \ud y_a \ud y^a \nn\\
&&+ \frac{K_1 (\r)}{\r} \left[\ud y_1 - \left((K'_1 (\r))^{-1}-1\right)\mbox{cosh}\w_2 ~\ud t'' \right.\nn\\
&&\left.\qquad\qquad\qquad- \left((K'_1 (\r))^{-1}-1\right)\mbox{sinh}\w_2 ~\ud y_2\right]^2 \nn \\
&&+ \frac{K_2 (\r)}{\r} \left[\ud y_2 - \left((K'_2 (\r))^{-1}-1\right) \ud t''\right]^2\,, \nn \\
f(\r) &=& 1 - m^{2\s} \r^\s\,, \nn\\
K_1(\r) &=& \left(1 + Q_1 \r^{\s}\right)\,,\qquad \left(K'_1(\r)\right)^{-1} = \left(1 - \bar{Q}_1 \r^{\s} K_1(\r)^{-1}\right)\,, \nn\\
K_2(\r) &=& \left(1 + Q_2 \r^{\s}K_1(\r)^{-1}\right)\,,\qquad \left(K'_2(\r)\right)^{-1} = \left(1 - \bar{Q}_2 \r^{\s} K_1(\r)^{-1}K_2(\r)^{-1}\right)\,, \nn\\
\eea
where 
\bea
&&Q_1 = m^{2 \s} \sinh^2 \omega_1\,, \qquad\qquad\ \ \ \ \ \bar{Q}_1 = m^{2 \s} \sinh \w_1 \cosh \omega_1\,, \nn\\
&&Q_2 = m^{2 \s} \sinh^2 \omega_2 \cosh^2 \w_1\,, \qquad \bar{Q}_2 = m^{2 \s} \sinh \w_2 \cosh \omega_2 \cosh^2 \w_1\,. \nn
\eea
We remove the wave by setting $\w_1=\w_2=0$, and recover the extremal limit by setting $m\to 0$ with $\w_1,\w_2\to\infty$ and $Q_1$, $Q_2$ finite.

Our aim is to perform a dimensional reduction as in section \ref{subsec:genred}, so we first rewrite the geometry (\ref{bbraneconfdt2}) to fit the structure of (\ref{redans}), or equivalently (\ref{redansfull}), and then insist that the coordinates $(y_1,y_2,y^a)$ are periodically identified with period $2 \pi R$. We then boost this resulting geometry along the non-compact boundary dimensions with boost parameter $\hat u_i$, satisfying $\eta_{ij}\hat u^i\hat u^j=-1$, where now $z^{i} = (t,z^r)$. We also wish to include two external, uniform gauge fields $\tilde{A}_{(0)i}^{(1)}\ud z^i$ and $\tilde{A}_{(0)i}^{(2)}\ud z^i$. We do so by performing coordinate transformations on $y_1$ and $y_2$, as $\ud y_1\to\ud \tilde{y}_1=\ud y_1 +\tilde{A}_{(0)i}^{(1)}\ud z^i$ and $\ud y_2\to\ud \tilde{y}_2=\ud y_2+\tilde{A}_{(0)i}^{(2)}\ud z^i$ respectively. All of these considerations ultimately yield
\bea
\ud s_{(2\s+1)}^2 &=& \frac{\ud\r^2}{4\r^2 f(\rho)} + \frac{1}{\rho} \left(\ud z^{i} \ud z_{i}\right) + \frac{1}{\rho} \left(1 - K_1(\rho)^{-1} K_2(\r)^{-1}f(\rho)\right) \hat u_{i}\hat u_{j} \ud z^{i} \ud z^{j} \nn\\
&&+ \frac{K_1 (\r)}{\r} \left[ \ud \tilde{y}_1 - \(\tilde{A}_{i(0)}^{(1)} + \left((K'_1 (\r))^{-1}-1\right)\mbox{cosh}\w_2 ~\hat u_{i} \right.\right. \nn\\
&& \left.\left. \qquad\qquad\qquad\qquad- \left((K'_1 (\r))^{-1}-1\right)\sinh\w_2 \tilde{A}_{i(0)}^{(2)} \)\ud z^i \right. \nn\\
&& \left. \qquad\qquad\qquad- \left((K'_1 (\r))^{-1}-1\right)\sinh\w_2~\ud \tilde{y}_2\right]^2 \nn\\
&&+ \frac{K_2 (\r)}{\r} \left[\ud \tilde{y}_2-\(\tilde{A}_{i(0)}^{(2)} + \left((K'_2 (\r))^{-1}-1\right)\hat{u}_i\)\ud z^i \right]^2 \nn\\
&& +\ \frac{1}{\rho} \ud y_a \ud y^a\,.
\eea

The next step towards deriving the universal hydrodynamics involves allowing the temperature, charge, fluid velocity and external gauge field to become position dependent, and correcting the metric at each order to satisfy the field equations. 

\subsubsection{Dimensional reduction}

The dimensional reduction involves comparing the metric with (\ref{redans}) and reading off the reduced metric, scalar fields, gauge fields and axion (we choose to read off $\phi_1$, $\phi_2$ and $\phi_3$ and then use (\ref{psizeta}) and (\ref{phixi}) to evaluate $\y$, $\z$, and $\x$, as opposed to reading these off (\ref{redansfull}), but the latter will yield the same answer). Since we are making contact with section \ref{subsec:hologdic}, in which the reduced metric is written in Fefferman-Graham form and all the fields are expanded using the Fefferman-Graham coordinate $\rho$ (which we call $\tilde{\r}$ in this section), we will also write all of our quantities of interest in Fefferman-Graham coordinates. This involves a redefinition of the radial coordinate $\r$ as per \cite{Kanitscheider:2009as}:
\be
\tilde{\r}(\r) = \(\frac{2}{1+\sqrt{1-m^{2\s}\r^{\s}}}\)^{2/\s}\r \qquad\Rightarrow\qquad \r(\tilde{\r})=\(1+\frac{m^{2\s}\tilde{\r}^{\s}}{4}\)^{-2/\s}\tilde{\r}\,.
\ee
The reduced metric is thus given by
\bea\label{Tmetriccalc}
\ud s_{(d+1)}^2 &=& \frac{\ud\r^2}{4\r^2 f(\rho)} + \frac{1}{\rho} \left(\ud z^{i} \ud z_{i}\right) + \frac{1}{\rho} \left(1 - K_1(\rho)^{-1}K_2(\rho)^{-1} f(\rho)\right) \hat{u}_{i} \hat{u}_{j} \ud z^{i} \ud z^{j}\nn\\
&=& \frac{\ud\tilde{\r}^2}{4\tilde{\r}^2} + \frac{1}{\tilde{\r}} \left( 1 + \frac{m^{2\s}\tilde{\r}^\s}{4} \right)^{\frac{2}{\s}} \ud z_i \ud z^i  \nn\\
&& + \frac{1}{\tilde{\r}} \left( 1 + \frac{m^{2\s}\tilde{\r}^\s}{4} \right)^{\frac{2}{\s}}
 \left[ 1 - K_1(\r(\tilde{\r}))^{-1}K_2(\r(\tilde{\r}))^{-1} f(\r(\tilde{\r})) \right] \hat{u}_i \hat{u}_j \ud z^i \ud z^j\,,\nn\\
 &&\mbox{\ \ } 
\eea
with the scalar fields being
\bea
e^{2 \phi_1} &=& \frac{K_1(\rho)}{\rho} = \frac{K_1(\r(\tilde{\r}))}{\tilde{\r}} \left( 1 + \frac{m^{2\s}\tilde{\r}^\s}{4} \right)^{\frac{2}{\s}} \,, \nn\\
e^{\frac{2 \phi_2}{(2 \s - d - 2)}} &=& \frac{1}{\rho} =\frac{1}{\tilde{\r}} \left( 1 + \frac{m^{2\s}\tilde{\r}^\s}{4} \right)^{\frac{2}{\s}} \,, \nn\\
e^{2 \phi_3} &=& \frac{K_2(\rho)}{\rho} = \frac{K_2(\r(\tilde{\r}))}{\tilde{\r}} \left( 1 + \frac{m^{2\s}\tilde{\r}^\s}{4} \right)^{\frac{2}{\s}}\,.
\eea
Rewriting the scalar fields in terms of $(\y,\z,\x)$ we obtain
\bea\label{ScalarFields}
e^{\y} &=& \frac{1}{\rho^{\sigma - d/2}} K_1(\rho)^{1/2}K_2(\rho)^{1/2} = \frac{K_1(\r(\tilde{\r}))^{\half}K_2(\r(\tilde{\r}))^{\half}}{\tilde{\r}^{\frac{2\s-d}{2}}} \left( 1 + \frac{m^{2\s} \tilde{\r}^\s}{4} \right)^{\frac{2\s-d}{\s}}\,, \nn\\
e^{\x} &=&  K_1(\r(\tilde{\r}))^{\frac{2\s-d-2}{2}}\,, \nn\\
e^{\z} &=& K_2(\r(\tilde{\r}))^{\frac{2\s-d-1}{2}}K_1(\r(\tilde{\r}))^{-\half}\,.
\eea
Furthermore, since
\be
e^{\frac{2\y}{2\s-d}} = \frac{1}{\tilde{\r}} e^{\frac{2\k}{2\s-d}} \,,\nn
\ee
we get that
\be
e^{\k} = K_1(\r(\tilde{\r}))^{\half}K_2(\r(\tilde{\r}))^{\half} \left( 1 + \frac{m^{2\s}\tilde{\r}^\s}{4} \right)^{\frac{2\s-d}{\s}}\,.
\ee
The gauge field $A^{(2)}$ is given by
\bea \label{ReducedGaugeField1}
A^{(2)} &=& \tilde{A}_{(0)i}^{(2)}\ud z^i +\left[(( K_2^{'}(\r(\tilde{\r})))^{-1} - 1)\hat{u}_i\right] \ud z^i\,,\nn\\
\eea
with
\be
A^{(1)} = \tilde{A}_{(0)i}^{(1)}\ud z^i +\left[(( K_1^{'}(\r))^{-1} - 1)\cosh\w_2~\hat{u}_i - (( K_1^{'}(\r))^{-1} - 1)\sinh\w_2~\tilde{A}_{(0)i}^{(2)}\right] \ud z^i\,, 
\ee
and the axion is
\be\label{axion}
A^{(0)}=\left((K'_1 (\r(\tilde{\r})))^{-1}-1)\right)\sinh\w_2\,.
\ee
The other gauge field we are actually interested in, and which sources $\hat{J}^{(1)}$ in the previous section (see (\ref{redforms}) and (\ref{Sren_check})) is
\bea\label{ReducedGaugeField2}
A^{(3)}&=&A^{(1)}+A^{(0)}A^{(2)} \nn\\
&=&\( \tilde{A}_{(0)i}^{(1)} +(( K_1^{'}(\r(\tilde{\r})))^{-1} - 1)\(\cosh\w_2 + \sinh\w_2(( K_2^{'}(\r(\tilde{\r})))^{-1} - 1)\)\hat{u}_i \)\ud z^i\,. \nn\\
&&\mbox{\ \ } 
\eea

\subsubsection{Expansion in Fefferman-Graham coordinates}

The next step is to expand the results in the previous section in the Fefferman-Graham radial co-ordinate $\tilde{\r}$, which gives us 
\bea
\label{expansions}
\k_{(0)} = 0 &;& \k_{(2\s)} = \half (Q_1 +Q_2) + \frac{(2\s-d)}{\s} \frac{m^{2\s}}{4}\,, \\
\z_{(0)} = 0 &;& \z_{(2\s)} = -\half Q_1+ \frac{(2\s-d-1)}{2} Q_2\,, \nn \\
\x_{(0)} = 0 &;& \x_{(2\s)} = \frac{(2\s-d-2)}{2} Q_1\,, \nn \\
A_{(0)}^{(0)}=0 &;& A_{(2\s)}^{(0)}=-\bar{Q}_1\sinh\w_2\,, \nn\\ 
A_{i(0)}^{(1)}=\tilde{A}_{i(0)}^{(1)}&;&A_{i(2\s)}^{(1)}=\bar{Q}_1\(\sinh\w_2 \tilde{A}_{(0)i}^{(2)}-\cosh\w_2 \hat{u}_i \) \,,\nn\\
A_{i(0)}^{(2)}=\tilde{A}_{i(0)}^{(2)}&;&A_{i(2\s)}^{(2)} = -\bar{Q}_2\hat{u}_i \,,\nn\\
A_{i(0)}^{(3)}=\tilde{A}_{i(0)}^{(1)}&;&A_{i(2\s)}^{(3)} = -\bar{Q}_1\cosh\w_2\hat{u}_i\,,  \nn\\
\tilde{g}_{(0)ij} = \eta_{ij} &;& \tilde{g}_{(2\s)ij} = \frac{m^{2\s}}{2\s}\eta_{ij} + \(Q_1 + Q_2 + m^{2\s}\)\hat{u}_i \hat{u}_j \,.\nn
\eea
We now have all the ingredients necessary to extract the expectation values of the dual operators from (\ref{vev}), and we obtain
\bea
\label{dualop}
\< \hat{T}_{ij} \>_d &=& L m^{2\s} \eta_{ij} + 2\s L(Q_1 +Q_2 + m^{2\s}) \hat{u}_i \hat{u}_j \nn\\
&=& L m^{2\s} \left(\eta_{ij} + 2\s\cosh^2 \omega_1\cosh^2 \omega_2~ \hat{u}_i \hat{u}_j \right) \,, \nn \\
\< \hat{J}_{i}^{(1)} \>_d &=& 2\s L \bar{Q}_1\cosh\w_2 ~\hat{u}_i \nn \\
&=& 2\s L m^{2\s} \sinh \omega_1 \cosh \omega_1\cosh\w_2~ \hat{u}_i \,, \nn \\
\< \hat{J}_{i}^{(2)} \>_d &=& 2\s L \bar{Q}_2 ~\hat{u}_i \nn \\
&=& 2\s L m^{2\s} \sinh \omega_2 \cosh \omega_2 \cosh^{2}\w_1~\hat{u}_i \,, \nn \\
\< {\cal O}_{1} \>_d &=& - Lm^{2\s}  - 2\s LQ_2 \nn \\
&=& -L m^{2\s} \left( 1 + 2\s \sinh^2 \omega_2 \cosh^2 \omega_1  \right)\,, \nn \\
\< {\cal O}_{2} \>_d &=& -L m^{2\s}\,,\nn\\
\< {\cal O}_{3} \>_d &=& - Lm^{2\s}  - 2\s LQ_1 \nn \\
&=& -L m^{2\s} \left( 1 + 2\s \sinh^2 \omega_1 \right)\,, \nn \\
\< {\cal O}_{4} \>_d &=& -2\s L \bar{Q}_1 \sinh\w_2 \nn\\
&=& -2\s L m^{2\s}\sinh\w_1\cosh\w_1\sinh\w_2\,,
\eea
and using (\ref{ops})
\bea
\<{\cal O}_{\y}\>_d &=& - Lm^{2\s} - \frac{2\s L}{(2\s-d)}\(Q_1 + Q_2\) \nn\\
&=& - Lm^{2\s} - \frac{2\s L m^{2\s}}{(2\s-d)}\(\sinh^2\w_1 + \sinh^2\w_2\cosh^2\w_1\)\,, \nn\\
\<{\cal O}_{\x}\>_d &=& -\frac{2\s L Q_1}{(2\s-d-1)} = -\frac{2\s L m^{2\s}}{(2\s-d-1)}\sinh^2\w_1\,,\nn\\
\<{\cal O}_{\z}\>_d &=& \frac{2\s L}{(2\s-d)}\(\frac{Q_1}{(2\s-d-1)} - Q_2\) \nn\\
&=& \frac{2\s L m^{2\s}}{(2\s-d)}\(\frac{\sinh^2\w_1}{(2\s-d-1)} - \sinh^2\w_2 \cosh^2\w_1\)\,, \nn\\
\<{\cal O}_{A^{(0)}}\>_d &=& 2\s L \bar{Q}_1\sinh\w_2 = 2\s L m^{2\s}\sinh\w_1\cosh\w_1\sinh\w_2\,.
\eea
Upon plugging the above into (\ref{dil_WI}), we see that the dilatation Ward identity is satisfied.

\subsubsection{Thermodynamic quantities}

Now, at thermal equilibrium
\bea\label{th1}
\<\hat{T}_{ij}\>_{d}=\hat{P}\eta_{ij}+(\hat{P}+\hat{\e})\hat{u}_i\hat{u}_j\,,\qquad \<\hat{J}^{(1)}_{i}\>_{d}=\hat{q}_{1}\hat{u}_{i}\,,\qquad \<\hat{J}^{(2)}_{i}\>_{d}=\hat{q}_{2}\hat{u}_{i} \,,
\eea
where $\hat \epsilon$ is the energy density, $\hat{q}_1$ and $\hat{q}_2$ the charge densities and $\hat P$ the pressure of the fluid dual to the reduced spacetime \eqref{Tmetriccalc}. Thus, from the expressions (\ref{dualop}) we can also read off the thermodynamic quantities,
\bea \label{thermo}
&&\hat \epsilon= L m^{2\s} (2 \s \cosh^2 \w_1 \cosh^2 \w_2 -1)\,, \nn\\
&&\hat{q}_1 \equiv 2\s L\bar{Q}_1\cosh\w_2 =  2\s L m^{2\s} \sinh \omega_1 \cosh \omega_1\cosh\w_2\,,\nn\\
&&\hat{q}_2 \equiv 2\s L\bar{Q}_2 =  2\s L m^{2\s} \sinh \omega_2 \cosh \omega_2\cosh^2\w_1\,,\nn\\
&&\hat P=L m^{2\s}\,.
\eea
In addition, from (\ref{ReducedGaugeField1}) and (\ref{ReducedGaugeField2}) we obtain that the chemical potentials\footnote{We may use regularity at the horizon, namely $\left . \hat{u}^i A_{i}^{(I)}\right |_{\r=m^{-2}} =0$ $(I=2,3)$, to relate the chemical potentials with the external gauge fields, respectively. However we choose not to do this, since we would like to keep the external gauge fields and chemical potentials as separate quantities when discussing the universal hydrodynamics in the next section. This allows us to make use of previously obtained results which also keep the two quantities distinct, especially the equations from which we extract transport coefficients in section \ref{subsec:univhydro_2}.} are equal to
\bea \label{chemical}
\hat \m_{1} &=& -\left(\left.\hat u^iA_i^{(3)}\right|_{\r=0}-\left.\hat u^iA_i^{(3)}\right|_{\r=m^{-2}}\right) = \frac{\tanh\w_1}{\cosh\w_2}\,, \nn\\
\hat \m_{2} &=& -\left(\left.\hat u^iA_i^{(2)}\right|_{\r=0}-\left.\hat u^iA_i^{(2)}\right|_{\r=m^{-2}}\right) = \tanh\w_2\,.
\eea
The thermodynamic identities
\be
\hat P+\hat \e = \hat T \hat s + \hat q_1 \hat \m_{1} + \hat q_2 \hat \m_{2}\,, \qquad \ud \hat P = \hat s \ud \hat T + \hat q_1 \ud \hat \m_{1} + \hat q_2 \ud \hat \m_{2}\,,
\ee
indeed hold under these thermodynamic values. 

Inverting the expressions in (\ref{thermo}) to express $m$, $\w_1$ and $\w_2$ in terms of $\hat \e$, $\hat{q}_1$ and $\hat{q}_2$, and then using the result in the last expression gives us the equation of state
\bea \label{pressure}
\hat P(\hat \e,\hat{q}_1,\hat{q}_2)&=&\frac{1}{2 \sigma-1} \left(\sqrt{(\hat \e (\s-1)-\hat{q}_1\hat{q}_2)^2+(\hat \e^2- (\hat{q}_1 +\hat{q}_2)^2) (2 \s-1)} \right. \nn\\
&& \left. \qquad\qquad - \hat \e (\s-1)+\hat{q}_1\hat{q}_2 \right)\,,
\eea
which we may use to evaluate the adiabatic speed of sound \cite{landau:1987:fm}, 
\be\label{sos}
\hat c_s^2\equiv\left.\frac{\partial \hat P}{\partial\hat \epsilon}\right|_{\hat s/\hat q_1,\ \hat s/\hat q_2} =\frac1{2(\s-1)\cosh^2\w_1 \cosh^2\w_2+1}\,,
\ee
where the ratios $\hat s/\hat q_1$ and $\hat s/\hat q_2$ are kept fixed, and $\hat{s}$ is the reduced entropy density given by 
\be\label{shat}
\hat s = 4 \pi L \cosh \omega_1 \cosh \omega_2 m^{2 \sigma-1}\,.
\ee
Analysing (\ref{pressure}) we see that if $\hat{q}_1,\hat{q}_2\to 0$, we get $\hat{P}=\hat{\e}/(2\s-1)$, the equation of state for non-conformal branes. Furthermore, in the extremal limit, where $\hat{\e}\to \left| \hat{q}_1+\hat{q}_2\right|$, $\hat{P}\to 0$ as expected, while above extremality  $\hat{\e}> \left| \hat{q}_1+\hat{q}_2\right|$ the expression under the square root is manifestly positive. Lastly, from (\ref{pressure}) and (\ref{thermo}) we see that  
\bea
m&=&\left[ \frac{1}{L(2 \sigma-1)} \left(\sqrt{(\hat \e (\s-1)-\hat{q}_1\hat{q}_2)^2+(\hat \e^2- (\hat{q}_1 +\hat{q}_2)^2) (2 \s-1)} \right. \right. \nn\\
&& \left. \left. \qquad\qquad - \hat \e (\s-1)+\hat{q}_1\hat{q}_2 \right)\right]^{1/2\s}\,.
\eea

The reduced temperature $\hat{T}$ is given by
\be \label{ReducedT}
\hat T = \frac{ m \sigma}{2 \pi \cosh \w_1\cosh \w_2}\,.
\ee

With the thermodynamic quantities in hand, we also notice that the expectation value of the scalar operator $\< {\cal O}_{\y} \>_d$ can be expressed in terms of the energy density and pressure as
\bea \label{dualop1}
\< {\cal O}_{\y} \>_d &=& \frac{1}{(2 \sigma -d)} \< {T}_{i}^i \>_d = \frac{1}{(2 \sigma -d)} \left [ (d-1) \hat P - \hat \ep \right ]\,,\\
\< {\cal O}_{\zeta} \>_d + \< {\cal O}_{\xi} \>_d &=& \frac{1}{(2 \sigma -d)} \left [ (2 \sigma -1) \hat P - \hat \ep \right ]\,, 
\eea
which shows that $\< {\cal O}_{\y} \>_d$ characterizes the deviation of the equation of state from conformality (as one would expect). The combination $\< {\cal O}_{\zeta} \>_d+ \< {\cal O}_{\xi} \>_d$ is zero in the uncharged case, which yields the equation of state of the non-conformal branes. This is also obtained by taking $q_1,q_2\to 0$ in (\ref{pressure}). 

\subsection{Universal hydrodynamics}\label{subsec:univhydro_2}

In this section we again turn to generalised dimensional reduction in order to obtain the universal hydrodynamics of the charged dilatonic solutions. We are interested in hydrodynamics at first-derivative order. Thus, in the upstairs picture we consider a conformal fluid on a curved manifold in the Landua-Lifshitz frame. Dimensionally reducing the upstairs energy-momentum tensor should in principle yield the transport coefficients of the lower-dimensional theory, but we find that the reduction does not leave us in the Landau-Lifshitz frame in the downstairs picture. Thus, we employ a frame-independent analysis to simplify things. This method uses the requirement that the divergence of the entropy current is positive semi-definite to write down equations from which we may extract the transport coefficients. With these transport coefficients in hand, we discuss various bounds which they might satisfy. Our analysis follows that in \cite{ouroldpaper}, except that now we are dealing with an additional charge in our system. Again, by setting $\w_1=0$ and $\w_2=\w$ in all results in this section, we recover the results in \cite{ouroldpaper}.

\subsubsection{Conformal fluid in higher dimension}

As in \cite{ouroldpaper}, we begin with a conformal fluid in $(2\s)$ dimensions on a curved manifold with metric $g\sub{0}_{\m\n}$, whose hydrodynamic energy-momentum tensor at first-derivative order is given by
\bea
\label{hydroconfTmn}
  \< T_{\mu \nu} \>_{2\s}&=& \<T_{\mu \nu}^{\rm eq}\>_{2\s} + \<T_{\mu \nu}^{\rm diss}\>_{2\s}\,, \\
  \< T_{\mu \nu}^{\rm eq}\>_{2\s}&=&P(g\sub{0}_{\mu \nu} + 2\s u_{\mu} u_{\nu})\,, \qquad
  \<T_{\mu \nu}^{\rm diss}\>_{2\s}= - 2\eta_{2\s} \s_{\mu \nu}\,, \nono \\
   \s_{\mu \nu} &=& P_{\mu}^{\kappa} P_{\nu}^{\lambda} \na_{(\kappa} u_{\lambda)}
   - \frac{1}{2\s-1} P_{\mu \nu} (\na \cdot u)\,, \qquad P_{\mu \nu} = g\sub{0}_{\mu \nu} + u_{\mu} u_{\nu}\,, \nono
\eea
where $T$, $u_{\mu}$ and $\eta_{2\s}$ denote the temperature, velocity and shear viscosity respectively of the fluid and $\na_{\mu}$ is the covariant derivative corresponding to the metric $g\sub{0}_{\mu \nu}$. The conservation of the energy-momentum tensor,
\be
\label{convcons}
   \na^{\mu} \<T_{\mu \nu}\>_{2\s} = 0\,,
\ee
determines the evolution of the fluid; note also that we are working in Landau-Lifshitz frame, where
\be
\label{LLcons}
u^{\mu}\<T^{\rm diss}_{\mu \nu}\>_{2\s}=0\,.
\ee

For later reference, note that for the AdS black brane,
\be \label{PressureAdSbbrane}
	P=L_{AdS} m^{2\s}\,, \qquad \eta_{2 \s}=\frac{s}{4\pi}=L_{AdS}m^{2\s-1}\,,
\ee
by \eqref{entropyAdSbbrane}.

As a first step towards writing down the universal hydrodynamics at first derivative order, we determine the reduced fluid velocity.

\subsubsection{Reduced fluid velocity}

We reduce the higher-dimensional fluid velocity $u^{\mu}$ by setting 
\be\label{conv}
u_a=0\,,\qquad u_{y_{1}} = \sinh\w_1\,, \qquad u_{y_{2}} = \cosh\w_1 \sinh\w_2\,,
\ee
with $(y_1,y_2,y^a)$ being compact dimensions, while ensuring that both
\be
u^{\mu} u_{\mu} =-1\,, \qquad u^{\mu} = g_{(0)}^{\mu \nu} u_{\mu}\,,
\ee
and
\be
\hat u^{i} \hat u_{i} =-1\,, \qquad \hat u^{i} = \h^{ij} \hat u_{j}\,,
\ee
hold. In the above, $\hat{u}^i$ is the lower-dimensional fluid velocity. The convention we have chosen in (\ref{conv}) serves as a link to the wave generating coordinate transformation of the previous subsection, and allows us to compare what we get \emph{via} (\ref{hydroconfTmn}), with what we obtained  previously in (\ref{dualop}). This requires us to first compute the boundary metric $g_{(0)\mu\nu}$, which we read off by using the reduction anzats (\ref{redansfull}) and the expansions of the fields (\ref{AdS_exp}), (\ref{AdS_exp_scalars}) and (\ref{AdS_exp_gauge}). This yields 
\bea \label{BoundaryMetric}
&&g_{(0)i j}=\h_{ij}+A_{(0) i}^{(1)}A_{(0) j}^{(1)} +A_{(0) i}^{(2)}A_{(0) j}^{(2)}\,, \qquad
g_{(0)i y_{1}}= -A_{(0) i}^{(1)}, \qquad g_{(0)i y_{2}}= -A_{(0) i}^{(2)} \,, \nn\\
&&\qquad\qquad\qquad g_{(0)y_{1}y_{1}}=1\,, \qquad  g_{(0)y_{2}y_{2}}=1\,, \qquad  g_{(0)y_{1}y_{2}}=0 \,, 
\eea
with the inverse metric given by
\bea \label{BoundaryInverseMetric}
&&\qquad\qquad g_{(0)}^{ij} = \eta^{ij}\,, \qquad g_{(0)}^{i y_{1}} = A_{(0)}^{(1)i}\,, \qquad g_{(0)}^{i y_{2}} = A_{(0)}^{(2)i}\,,\nn\\
&&g_{(0)}^{y_{1}y_{1}} =1+\eta^{ij}A_{(0) i}^{(1)}A_{(0) j}^{(1)}\,,\qquad g_{(0)}^{y_{2}y_{2}} =1+\eta^{ij}A_{(0) i}^{(2)}A_{(0) j}^{(2)}\,, \nn\\
&&\qquad\qquad\qquad\qquad g_{(0)}^{y_{1}y_{2}} =\eta^{ij}A_{(0) i}^{(1)}A_{(0) j}^{(2)}\,. 
\eea
For simplicity, and to make a connection to the case of the AdS black brane (\ref{expansions}), we have set $\k_{(0)}=\z_{(0)}=\x_{(0)}=A_{(0)}^{(0)}=0$ in the above, and in what follows. Note that the reduced boundary metric is the Minkowski metric.

With all this in mind, the reduction yields
\bea \label{u_red} 
u_{i} &=& \cosh \w_1\cosh\w_2\hat u_i-\sinh\w_1 A_{(0)i}^{(1)}-\cosh\w_1\sinh\w_2 A_{(0)i}^{(2)}\,,\nn\\
u_{y_{1}} &=& \sinh\w_1\,,  \nn\\
u_{y_{2}} &=& \cosh\w_1 \sinh\w_2 \,, \nn\\
u^{i} &=& \cosh\w_1\cosh\w_2 \hat u^i \,, \nn\\
u^{y_{1}}&=&\sinh\w_1+\cosh\w_1\cosh\w_2~\hat u\cdot A_{(0)}^{(1)}\,, \nn\\
u^{y_{2}}&=&\cosh\w_1\sinh\w_2+\cosh\w_1\cosh\w_2~\hat u\cdot A_{(0)}^{(2)}\,. 
\eea

\subsubsection{Equilibrium quantities}

We may now insert the values (\ref{u_red}) into (\ref{hydroconfTmn}), and using (\ref{redforms}) obtain the equilibrium (zero-derivative order) values for the stress-energy tensor, currents and operators:
\bea
\label{dualop2}
\< \hat{T}^{\rm eq}_{ij} \>_d &=& \hat P \left[\eta_{ij} + 2\s\left(u_i+u_{y_{1}}A_{(0)i}^{(1)}+u_{y_{2}}A_{(0)i}^{(2)}\right)\left(u_j+u_{y_{1}}A_{(0)j}^{(1)}+u_{y_{2}}A_{(0)j}^{(2)}\right)\right]\,, \nn \\
\< \hat{J}_{i}^{(1)\rm eq} \>_d  &=& 2\s\hat Pu_{y_{1}}\left(u_i+u_{y_{1}}A_{(0)i}^{(1)}+u_{y_{2}}A_{(0)i}^{(2)}\right)\,, \nn \\
\< \hat{J}_{i}^{(2)\rm eq} \>_d  &=& 2\s\hat Pu_{y_{2}}\left(u_i+u_{y_{1}}A_{(0)i}^{(1)}+u_{y_{2}}A_{(0)i}^{(2)}\right)\,, \nn \\
\< {\cal O}^{\rm eq}_{1} \>_d &=& - \hat P\left(1+2\s u_{y_{2}}^{2}\right)\,, \nn \\
\< {\cal O}^{\rm eq}_{2} \>_d\d_{ab} &=& - \hat P\left(\d_{ab}+2\s u_au_b\right)\,, \nn\\
\< {\cal O}^{\rm eq}_{3} \>_d &=& - \hat P\left(1+2\s u_{y_{1}}^{2}\right)\,, \nn \\
\< {\cal O}^{\rm eq}_{4} \>_d &=& - 2\s\hat P u_{y_{1}}u_{y_{2}}\,, \nn 
\eea
which yields
\bea
\label{dualop3}
\< \hat{T}^{\rm eq}_{ij} \>_d &=& \hat P \left(\eta_{ij} + 2\s\cosh^2\w_1\cosh^2\w_2\hat u_i\hat u_j\right)\,,  \\
\< \hat{J}_{i}^{(1)\rm eq} \>_d  &=& 2\s\hat P\sinh\w_1\cosh\w_1\cosh\w_2\hat u_i\,,  \\
\< \hat{J}_{i}^{(2)\rm eq} \>_d  &=& 2\s\hat P\cosh^2\w_1\cosh\w_2\sinh\w_2\hat u_i\,,  \\
\< {\cal O}^{\rm eq}_{1} \>_d &=& - \hat P\left(1+2\s \cosh^2\w_1\sinh^{2}\w_2\right)\,,  \\
\< {\cal O}^{\rm eq}_{2} \>_d &=& - \hat P\,,\\
\< {\cal O}^{\rm eq}_{3} \>_d &=& - \hat P\left(1+2\s \sinh^{2}\w_1\right)\,,  \\
\< {\cal O}^{\rm eq}_{4} \>_d &=& -2\s\hat P\sinh\w_1\cosh\w_1\sinh\w_2 \,, 
\eea
with
\bea\label{dualop3_1}
\< {\cal O}^{\rm eq}_{\y} \>_d &=& - \frac{\hat P}{(2\s-d)}\left(2\s\cosh^2\w_1\cosh^2\w_2-d\right)\,, \\
\< {\cal O}^{\rm eq}_{\x} \>_d &=& - \frac{2\s\hat P}{(2\s-d-1)}\sinh^2\w_1\,,\\
\< {\cal O}^{\rm eq}_{\z} \>_d &=&  \frac{2\s\hat P}{(2\s-d)(2\s-d-1)}\left(\sinh^2\w_1-(2\s-d-1)\cosh^2\w_1\sinh^2\w_2\right)\,,\nn\\
&&\mbox{\ \ } \\
\< {\cal O}^{\rm eq}_{A_{(0)}^{(1)}} \>_d &=& 2\s\hat{P}\sinh\w_1\cosh\w_1\sinh\w_2 \,.
\eea
Using (\ref{th1}), we can read off the equilibrium quantities
\bea
&&\qquad\qquad\hat P =\frac{L}{L_{AdS}}P\,,\qquad \hat \e = \left(2\s\cosh^2\w_1\cosh^2\w_2-1\right)\hat P\,,\nn\\
&&\hat q_1=2\s\hat P\sinh\w_1\cosh\w_1\cosh\w_2\,, \qquad \hat q_2=2\s\hat P\cosh\w_2\sinh\w_2\cosh^2\w_1\,. \nn\\
&&\mbox{\ \ } \label{eqmquants}
\eea
Using (\ref{PressureAdSbbrane}) for the pressure density of the AdS black brane recovers (\ref{thermo}) as well as the dual operators in \eqref{dualop}. 

\subsubsection{Dissipative extension and transport coefficient formulae}

We now move on to the dissipative part. We wish to evaluate the transport coefficients characterizing the reduced dual field theory. Since the upstairs picture comprises a conformal fluid in $(2\s)$ dimensions governed by (\ref{hydroconfTmn}), (\ref{convcons}) and (\ref{LLcons}), we see that the only transport coefficient in the upstairs picture is the shear viscosity $\eta_{2\s}$. In the downstairs picture we will also end up with a shear viscosity $\hat{\eta}$, but also bulk viscosity $\hat{\z}_s$ and heat conductivity $\hat{\k}_T$. We may obtain these transport coefficients by dimensionally reducing $\<T_{\mu \nu}^{\rm diss}\>_{2\s}$ in (\ref{hydroconfTmn}) using $u^{\mu} = (u^i, 0, u^{y_{1}},u^{y_{2}})$ from (\ref{u_red}). However, we find that reducing the Landau-Lifshitz frame condition (\ref{LLcons}) doesn't result in the reduced frame also being the Landau-Lifshitz frame. More precisely, as a first step in reducing $u^{\mu}\<T_{\mu\nu}^{\rm diss}\>_{2\s} = 0 $ we write
\bea
&\mbox{$\nu = j:$}&\qquad u^{i}\<T_{ij}^{\rm diss}\>_{2\s} + u^{y_1}\<T_{y_1 j}^{\rm diss}\>_{2\s} + u^{y_2}\<T_{y_2 j}^{\rm diss}\>_{2\s} = 0\,,\label{nuisj}\\
&\mbox{$\nu = y_1:$}&\qquad u^{i}\<T_{iy_1}^{\rm diss}\>_{2\s} + u^{y_1}\<T_{y_1 y_1}^{\rm diss}\>_{2\s} + u^{y_2}\<T_{y_2 y_1}^{\rm diss}\>_{2\s} = 0 \,,\label{nuisy1}\\
&\mbox{$\nu = y_2:$}&\qquad u^{i}\<T_{i y_2}^{\rm diss}\>_{2\s} + u^{y_1}\<T_{y_1 y_2}^{\rm diss}\>_{2\s} + u^{y_2}\<T_{y_2 y_2}^{\rm diss}\>_{2\s} = 0\,.\label{nuisy2}
\eea
Using (\ref{redforms}) these become 
\bea
\label{LLconsd}
\hat{u}^i \< \hat{T}_{ij}^{\rm diss} \>_d &=& -\frac{\tanh\w_1}{\cosh\w_2} \< \hat{J}_j^{(1)\rm diss}\> -\tanh\w_2 \< \hat{J}_j^{(2)\rm diss}  \>_d\,,\nn\\
\hat{u}^i \< \hat{J}_i^{(1)\rm diss} \>_d &=& \frac{\tanh\w_1}{\cosh\w_2} \< \cO_3^{\rm diss} \>_d + \tanh\w_2 \< \cO_4^{\rm diss} \>_d\,,  \nono \\
\hat{u}^i \< \hat{J}_i^{(2)\rm diss} \>_d &=& \frac{\tanh\w_1}{\cosh\w_2} \< \cO_4^{\rm diss} \>_d + \tanh\w_2 \< \cO_1^{\rm diss} \>_d\,.  \nono \\
\eea
More precisely, (\ref{nuisj}) becomes
\bea
&&\hat{u}^{i}\<\hat{T}_{ij}^{\rm diss}\>_d - \hat{u}^{i}A_{j(0)}^{(1)}\<\hat{J}_{i}^{(1)\rm diss}\>_d - \hat{u}^{i}A_{j(0)}^{(2)}\<\hat{J}_{i}^{(2)\rm diss}\>_d \nn\\
&& + \frac{\tanh\w_1}{\cosh\w_2}\left[ \<\hat{J}_{j}^{(1)\rm diss}\>_d + A_{j(0)}^{(1)}\<{\cal O}_{3}^{\rm diss}\>_d  + A_{j(0)}^{(2)}\<{\cal O}_{4}^{\rm diss}\>_d\right] \nn\\
&& + \tanh\w_2\left[ \<\hat{J}_{j}^{(2)\rm diss}\>_d + A_{j(0)}^{(1)}\<{\cal O}_{4}^{\rm diss}\>_d  + A_{j(0)}^{(2)}\<{\cal O}_{1}^{\rm diss}\>_d\right] = 0\,, \label{nuisj_1}
\eea
which yields the first expression in (\ref{LLconsd}) once the last two expressions in (\ref{LLconsd}) are substituted into it.

It is clear that (\ref{LLconsd}) does not represent a Landau-Lifshitz frame in the reduced theory. As a result, we will use the frame-independent analysis developed in \cite{Bhattacharya:2011tr}, which relies on ensuring that the divergence of the entropy current is positive semi-definite. In this case, starting from the $(2\s+1)$-dimensional entropy current in Landau-Lifshitz frame
\be \label{EntropyCurrent}
	\<J^\mu_s\>_{2\s}=su^\mu\,,
\ee
obeying divergence relation 
\be \label{DivEntropyCurrent}
	\nabla_\mu\<J^\mu_s\>_{2\s} = -\nabla_\mu\left(\frac{u_\nu}T\right)\<T^{\mu\nu}_{\rm diss}\>_{2\s}=-\frac1T\s_{\mu\nu}\<T^{\mu\nu}_{\rm diss}\>_{2\s}\,,
\ee
upon reduction we obtain
\be \label{ReducedEntropyCurrent}
	\<J^i_s\>_{d}=\hat s \hat u^i\,,\qquad \hat s = \frac{L\cosh\w_1\cosh\w_2}{L_{AdS}}s\,,
\ee
with 
\bea
\label{ReducedDivEntropyCurrent}
\nabla_i\<J^i_s\>_{d} &=& -\partial_i\left(\frac{\hat u_j}{\hat T}\right)\<\hat{T}^{ij}_{\rm diss}\>_{d} - \left[\partial_i\frac{\hat\mu_{1}}{\hat T}-\frac{\hat u_k}{\hat T}F_{(0)\ \ i}^{(1)k}\right]\<\hat{J}^{(1)i}_{\rm diss}\>_d \nn\\
&& \qquad\qquad\qquad\qquad\  - \left[\partial_i\frac{\hat\mu_{2}}{\hat T}-\frac{\hat u_k}{\hat T}F_{(0)\ \ i}^{(2)k}\right]\<\hat{J}^{(2)i}_{\rm diss}\>_d \,.\nn\\
&=& -\frac{(\partial\cdot \hat{u})}{\hat{T}}\left[ \frac{\<\hat{T}_{\rm diss}^{ij}\>\hat{P}_{ij}}{(d-1)} - \left . \(\frac{\partial \hat{P}}{\partial\hat{\e}}\)\right|_{\hat{q}_1,\hat{q}_2} \<\hat{T}_{\rm diss}^{ij}\>\hat{u}_{i}\hat{u}_{j} + \left . \(\frac{\partial \hat{P}}{\partial \hat{q}_1}\) \right|_{\hat{\e},\hat{q}_2} \<\hat{J}_{\rm diss}^{(1)i}\>\hat{u}_{i} \right.\nn\\
&&\qquad\qquad\qquad \left. + \left. \(\frac{\partial \hat{P}}{\partial \hat{q}_2}\)\right|_{\hat{\e},\hat{q}_1}\<\hat{J}_{\rm diss}^{(2)i}\>\hat{u}_{i}\right] - \frac{\<\hat{T}_{\rm diss}^{ij}\>\hat{\s}_{ij}}{\hat{T}}\nn\\
&&+\ V_{1i}^{(1)}\left[ \<\hat{J}_{\rm diss}^{(1)i}\> + \(\frac{\hat{q}_1}{\hat{P} + \hat{\e}}\) \<\hat{T}_{\rm diss}^{ij}\>\hat{u}_{j} \right]\nn\\
&&+\ V_{1i}^{(2)}\left[ \<\hat{J}_{\rm diss}^{(2)i}\> + \(\frac{\hat{q}_2}{\hat{P} + \hat{\e}}\) \<\hat{T}_{\rm diss}^{ij}\>\hat{u}_{j} \right]\,,
\eea
where, for $I=1,2$,
\be\label{Veqn}
V_{1i}^{(I)} = -\(P_{ij}\partial^{j}\frac{\hat{\mu}_I}{\hat{T}}+\frac{F^{(I)}_{(0)ij}\hat{u}^{j}}{\hat{T}}\)\,.
\ee
We obtain the last equality in (\ref{ReducedDivEntropyCurrent}) above by following the logic outlined in \cite{Bhattacharya:2011tr}, but extended to include an extra charge (in \cite{Bhattacharya:2011tr} they deal with a singly charged system). For more details on how we used the method of \cite{Bhattacharya:2011tr} to derive the above, and hence the following equations, refer to appendix \ref{app:transcoeffs2}. Requiring (\ref{ReducedDivEntropyCurrent}) to be positive semi-definite leads to the formulae
\bea
&&\hat{P}^i_k \hat{P}^j_l \<\hat{T}^{\rm diss}_{ij}\>_d
- \frac{1}{d -1} \hat{P}_{kl} \hat{P}^{ij} \<\hat{T}^{\rm diss}_{ij}\>_d
= - 2 \hat \eta \hat{\s}_{kl} \,, \label{TensorInv}\\
&&\hat{P}^j_i \left(\<\hat{J}^{(1)\rm diss}_j\>_d + \frac{\hat q_1}{\hat \e + \hat P} \hat{u}^k \< \hat{T}_{kj}^{\rm diss} \>_d \right) = -\hat \kappa_{11} \left(\hat{P}_{ij} \partial^j \frac{\hat \mu_{1}}{\hat T} + \frac{F_{(0)ij}^{(1)} \hat{u}^j}{\hat T}\right) \nn\\
&& \qquad\qquad\qquad\qquad\qquad\qquad\qquad\qquad\ \ -\hat{\kappa}_{12}\left(\hat{P}_{ij} \partial^j \frac{\hat \mu_{2}}{\hat T} + \frac{F_{(0)ij}^{(2)} \hat{u}^j}{\hat T}\right)\,, \label{VectorInv2}\\
&&\hat{P}^j_i \left(\<\hat{J}^{(2)\rm diss}_j\>_d + \frac{\hat q_2}{\hat \e + \hat P} \hat{u}^k \< \hat{T}_{kj}^{\rm diss} \>_d \right) = -\hat \kappa_{22} \left(\hat{P}_{ij} \partial^j \frac{\hat \mu_{2}}{\hat T} + \frac{F_{(0)ij}^{(2)} \hat{u}^j}{\hat T}\right)\nn\\
&& \qquad\qquad\qquad\qquad\qquad\qquad\qquad\qquad\ \ -\hat{\kappa}_{21}\left(\hat{P}_{ij} \partial^j \frac{\hat \mu_{1}}{\hat T} + \frac{F_{(0)ij}^{(1)} \hat{u}^j}{\hat T}\right)\,, \label{VectorInv}\\
&&\frac{\hat{P}^{ij} \<\hat{T}^{\rm diss}_{ij}\>_d}{d -1} - \frac{\partial \hat P}{\partial \hat \e} \hat{u}^i \hat{u}^j \< \hat{T}_{ij}^{\rm diss} \>_d +
\frac{\partial \hat P}{\partial \hat q_1} \hat{u}^i \< \hat{J}_i^{(1)\rm diss} \>_d +
\frac{\partial \hat P}{\partial \hat q_2} \hat{u}^i \<\hat{J}_i^{(2)\rm diss} \>_d =
- \hat \zeta_s \partial_i \hat{u}^i\,. \label{ScalarInv}\nn\\
&&\mbox{\ \ } 
\eea
In the above $\hat\eta\geq 0$, $\hat\z_s\geq 0$, while $\hat{\k}_{11}>0$, $\hat{\k}_{22}>0$, with $\det \hat{\k} =\(\hat{\k}_{11}\hat{\k}_{22}-\(\hat{\k}_{12}\)^2\)>0$. We may obtain these latter conditions involving $\hat{\kappa}_{ij}$, $(i,j=1,2)$ by noting that if we substitute (\ref{VectorInv2}) and (\ref{VectorInv}) into the last two lines of (\ref{ReducedDivEntropyCurrent}), we get
\bea
&&V_1^{(1)}\(\hat{\kappa}_{11}V_1^{(1)} + \hat{\kappa}_{12}V_1^{(2)}\) + V_1^{(2)}\(\hat{\kappa}_{21}V_1^{(1)}+\hat{\kappa}_{22}V_1^{(2)}\) \nn\\
&&=\hat{\kappa}_{11}\(V_1^{(1)}\)^2 + 2\hat{\kappa}_{12}V_1^{(1)}V_1^{(2)} + \hat{\kappa}_{22}\(V_1^{(2)}\)^2\,,
\eea
where $\hat{\kappa}_{12}=\hat{\kappa}_{21}$. The quadratic polynomial above is positive for all $V_1^{(1)},V_1^{(2)}$ if it is positive somewhere and has no real roots. Now, if $V_1^{(1)}>>V_1^{(2)}$, the polynomial is positive if $\hat{\kappa}_{11}>0$. No real roots implies
\be
(2\hat{\kappa}_{12})^2 - 4\hat{\kappa}_{11}\hat{\kappa}_{22}<0 \Rightarrow \det\hat{\k} >0\,,
\ee
which then implies that $\hat{\kappa}_{22}>0$. Similarly, if $V_1^{(2)}>>V_1^{(1)}$ the polynomial is positive if $\hat{\kappa}_{22}>0$, such that $\det\hat{\k}>0\Rightarrow\hat{\k}_{11}>0$.

Now applying the reduced conditions (\ref{LLconsd}) on the equations (\ref{VectorInv2}), (\ref{VectorInv}) and (\ref{ScalarInv}) yields
\bea
&&\hat{P}^{ij} \<\hat{J}^{(1)\rm diss}_j\>_d \left(1 - \frac{\hat q_1}{\hat \e + \hat P}\frac{\tanh\w_1}{\cosh\w_2} \right) - \hat{P}^{ij} \<\hat{J}^{(2)\rm diss}_j\>_d \left( \frac{\hat q_1}{\hat \e + \hat P}\tanh\w_2 \right) \nn\\
&&\qquad= -\hat \kappa_{11} \left(\hat{P}^{ij} \partial_j \frac{\hat \mu_{1}}{\hat T} + \frac{F^{(1)ij}_{(0)} \hat{u}_j}{\hat T}\right) -\hat{\kappa}_{12}\left(\hat{P}^{ij} \partial_j \frac{\hat \mu_{2}}{\hat T} + \frac{F_{(0)}^{(2)ij} \hat{u}_j}{\hat T}\right) \,, \label{BulkViscosityEqn_1} \\
&&\hat{P}^{ij} \<\hat{J}^{(2)\rm diss}_j\>_d \left(1 - \frac{\hat q_2}{\hat \e + \hat P}\tanh\w_2 \right) - \hat{P}^{ij} \<\hat{J}^{(1)\rm diss}_j\>_d \left( \frac{\hat q_2}{\hat \e + \hat P}\frac{\tanh\w_1}{\cosh\w_2} \right) \nn\\
&&\qquad= -\hat \kappa_{22} \left(\hat{P}^{ij} \partial_j \frac{\hat \mu_{2}}{\hat T} + \frac{F^{(2)ij}_{(0)} \hat{u}_j}{\hat T}\right)-\hat{\kappa}_{21}\left(\hat{P}^{ij} \partial_j \frac{\hat \mu_{1}}{\hat T} + \frac{F_{(0)}^{(1)ij} \hat{u}_j}{\hat T}\right)\,, \label{BulkViscosityEqn_2} \\
&&\frac{\hat{P}^{ij} \<\hat{T}^{\rm diss}_{ij}\>_d}{(d -1)} +\frac{\tanh\w_1}{\cosh\w_2}\left( \frac{\tanh\w_1}{\cosh\w_2}\frac{\partial \hat P}{\partial \hat \e}   +\frac{\partial \hat P}{\partial \hat q_1} \right) \< \cO_3^{\rm diss} \>_d \nn\\
&& \qquad+\tanh\w_2\left( \tanh\w_2 \frac{\partial \hat P}{\partial \hat \e} +\frac{\partial \hat P}{\partial \hat q_2} \right) \< \cO_1^{\rm diss} \>_d \nn\\
&&\qquad+\(\frac{2\tanh\w_1\tanh\w_2}{\cosh\w_2}\frac{\partial \hat P}{\partial \hat \e}+\tanh\w_2\frac{\partial \hat P}{\partial \hat q_1}+\frac{\tanh\w_1}{\cosh\w_2}\frac{\partial \hat P}{\partial \hat q_2} \) \< \cO_4^{\rm diss} \>_d \nn\\
&&\qquad\qquad\qquad\qquad\qquad\qquad\qquad\qquad\qquad\qquad\qquad= - \hat \zeta_s \partial_i \hat{u}^i\,.\label{BulkViscosityEqn} 
\eea
From these formulae we may thus extract the reduced transport coefficients $\hat{\eta}$, $\hat{\z}_s$ and the heat conductivity matrix $\hat{\k}_{ij}$, for $i,j=1,2$. The details of this are given in appendix \ref{app:comptranscoeff}. We obtain
\bea
\hat\eta&=&\eta_d\cosh\omega_1\cosh\w_2 = L m^{2\s-1}\cosh\omega_1\cosh\w_2\label{ShearViscosity}\\
&=& L\left(\frac{2\pi\hat T}{\sigma}\right)^{2\s-1}\left(1-\hat\mu_{1}^2 - \hat\mu_{2}^2\right)^{-\sigma}\,, \label{ShearViscosity2}\\
\hat\kappa_{11}&=& \eta_d\frac{\s m}{2\pi}\(1-\frac{\sinh^2\w_1}{\cosh^2\w_1\cosh^2\w_2}\)=\frac{\s Lm^{2\s}}{2\pi}\(1-\frac{\sinh^2\w_1}{\cosh^2\w_1\cosh^2\w_2}\)\label{HeatConductivity_1}\nn\\
&&\mbox{\ }\\
&=&\frac{\s L}{2\pi}\left(\frac{2\pi\hat T}{\sigma}\right)^{2\s}\left(1-\hat\mu_{1}^2 - \hat\mu_{2}^2\right)^{-\sigma}(1-\hat\mu_{1}^2)\,,\label{HeatConductivity11}\\
\hat\kappa_{22}&=&\eta_d\frac{\s m}{2\pi}\frac{1}{\cosh^2\w_2}=\frac{\s Lm^{2\s}}{2\pi}\frac{1}{\cosh^2\w_2}\label{HeatConductivity_2}\\
&=&\frac{\s L}{2\pi}\left(\frac{2\pi\hat T}{\sigma}\right)^{2\s}\left(1-\hat\mu_{1}^2 - \hat\mu_{2}^2\right)^{-\sigma}(1-\hat\mu_{2}^2)\,,\label{HeatConductivity22}\\
\hat{\kappa}_{12}&=&  \hat{\kappa}_{21} = -\eta_d\frac{\s m}{2\pi}\frac{\sinh\w_1\sinh\w_2}{\cosh\w_1\cosh^2\w_2}=-\frac{\s Lm^{2\s}}{2\pi}\frac{\sinh\w_1\sinh\w_2}{\cosh\w_1\cosh^2\w_2}\label{HeatConductivity_3}\\
&=& - \frac{\s L}{2\pi}\left(\frac{2\pi\hat T}{\sigma}\right)^{2\s}\left(1-\hat\mu_{1}^2 - \hat\mu_{2}^2\right)^{-\sigma}\hat\mu_{1}\hat\mu_{2}\,,\label{HeatConductivity2}\\
\hat\z_s &=& 2\eta_d \cosh\w_1\cosh\w_2 \nn\\
&&\times\left[ \frac{1}{(d-1)}-\frac{(2(\s-1)\cosh^4\w_1\cosh^4\w_2+2\cosh^2\w_1\cosh^2\w_2-1)}{(2(\s-1)\cosh^2\w_1\cosh^2\w_2+1)^2} \right]\label{BulkViscosity} \nn\\  
&&\mbox{\ } \\
&=&2\hat\eta\left[\frac{1}{d-1}-\frac{\(2\s-1- (\hat\mu_{1}^2 + \hat\mu_{2}^2)^2\)}{\(2\s-1- \hat\mu_{1}^2 - \hat\mu_{2}^2\)^2}\right]\,,\label{BulkViscosity2}  
\eea
where $\eta_d$ is the shear viscosity of the reduced uncharged case,
\be \label{ShearNeutral}
	\eta_d = \frac{L}{L_{AdS}} \eta_{2 \s}= L m^{2\s-1}\,.
\ee
In the above we have also re-expressed all the transport coefficients for the reduced AdS black brane  in terms of the temperature and chemical potentials. Note also that the conditions required for our transport coefficients are indeed satisfied: $\hat\z_s\geq 0$, $\hat\eta\geq 0$, while $\hat{\k}_{11}>0$, $\hat{\k}_{22}>0$, with $\hat{\k}_{11}\hat{\k}_{22}-\(\hat{\k}_{12}\)^2 = \(\frac{\s L m^{2\s}}{2\pi}\)^2 / (\cosh^2\w_1\cosh^2\w_2)>0$. We may use the heat conductivity matrix $\hat \kappa_{ij}$ for $i,j=1,2$ to obtain the heat conductivity using \cite{SonStarinets:2006,Jain:2010},
\bea\label{newkappa}
\hat{\k}_{T} \equiv \(\frac{\hat{\e}+\hat{P}}{\hat{T}}\)^2\frac{1}{\displaystyle\sum\limits_{i,j=1}^{2}\hat{q}_i\hat{\k}_{ij}^{-1}\hat{q}_j} &=& \frac{2\pi L m^{2\s-2}}{\s}\frac{\cosh^2\w_1 \cosh^2\w_2}{\(\cosh^2\w_1 \cosh^2\w_2 -1\)} \\
&=& \frac{2\pi L}{\s}\(\frac{2\pi\hat{T}}{\s}\)^{2\s-2}\frac{\(1-\hat\mu_{1}^2 - \hat\mu_{2}^2\)^{1-\s}}{\hat\mu_{1}^2 + \hat\mu_{2}^2}\,.
\eea

A recent formula for $\hat{\z}_s/\hat{\eta}$ first developed in \cite{Eling:2011ms}, allows us to check our value for the shear to bulk viscosity ratio formed from (\ref{BulkViscosity}) and (\ref{ShearViscosity}) 
\be\label{ratio_1}
\frac{\hat{\z}_s}{\hat{\eta}} = 2\left[ \frac{1}{(d-1)}-\frac{(2(\s-1)\cosh^4\w_1\cosh^4\w_2+2\cosh^2\w_1\cosh^2\w_2-1)}{(2(\s-1)\cosh^2\w_1\cosh^2\w_2+1)^2} \right]\,.
\ee
The formula is quoted in \cite{Eling:2011ms} as
\be
    \frac{\hat\z_s}{\hat\h} = \sum_I\left(\hat s\frac{\ud\phi_h^I}{\ud\hat s}+\hat q_a\frac{\ud\phi_h^I}{\ud\hat q_a}\right)^2\,,
    \label{ElingOzFormula2}
\ee
where $\hat q_a$ are conserved charge densities and $\phi^I_h$ are a collection of scalar fields, evaluated at the event horizon, and the formula is valid in the Einstein frame where the entropy density $\hat{s}$ is given by the quarter of the horizon area. This formula was derived for an action in which the scalar fields $\phi^I$ are canonically normalised. However,  in our case we need to adjust this formula slightly to account for the non-standard normalisation of the axion kinetic term in our action (\ref{finact}). Following the same procedure in appendix \ref{app:derivratio} as outlined in \cite{Eling:2011ms}, we arrive at the following formula:
\be
\frac{\hat{\z}_s}{\hat{\eta}}=\Omega_{IJ}\(\hat{s}\frac{\ud\phi_{h}^{I}}{\ud\hat{s}} + \hat{q}^a\frac{\ud\phi_{h}^{I}}{\ud\hat{q}^a}\)\(\hat{s}\frac{\ud\phi_{h}^{J}}{\ud\hat{s}} + \hat{q}^a\frac{\ud\phi_{h}^{J}}{\ud\hat{q}^a}\)\,,
    \label{ElingOzFormula1}
\ee
where $\phi^I = \{\bar{\y},\bar{\z},\bar{\x},A^{(0)}\}$, and $\Omega_{IJ}=\mbox{diag}\{1,1,1,\(\frac{\cosh\w_1}{\cosh\w_2}\)^2\}$. Note that $\Omega_{IJ}=\mbox{diag}\{1,1,...,1\}$ reproduces (\ref{ElingOzFormula2}), and corresponds to all the scalar kinetic terms being canonically normalised as done in \cite{Eling:2011ms}. The entropy and charge density in the Einstein frame are still given by \eqref{shat} and \eqref{thermo}, the scalars we use in (\ref{ElingOzFormula1}) are obtained from (\ref{ScalarFields}) but further normalized as in (\ref{NormScalars}), and the axion is obtained from (\ref{axion}). We provide the details in re-evaluating $\hat{\z}_s/\hat{\eta}$ using formula (\ref{ElingOzFormula1}) in appendix \ref{app:evalratio}.

The DC conductivity can be deduced using the Wiedemann-Franz law:
\bea
\hat{\s}_{DC} =c_1 \frac{\hat{\k}_T}{\hat{T}} &=&c_1 \frac{4\pi^2 L m^{2\s-3}}{\s^2}\frac{\cosh^3\w_1\cosh^3\w_2}{(\cosh^2\w_1\cosh^2\w_2 -1)} \nn\\
&=&c_1 \(\frac{2\pi}{\s}\)^2 L\(\frac{2\pi\hat{T}}{\s}\)^{2\s-3}\frac{\(1-\hat\mu_{1}^2 - \hat\mu_{2}^2\)^{1-\s}}{\hat\mu_{1}^2 + \hat\mu_{2}^2}\,,
\eea
where $c_1$ is a proportionality constant. Furthermore we note that the ratio of thermal conductivity and shear viscosity obeys a simple relation similar to the Wiedemann-Franz law, even in the presence of multiple chemical potential:
\be
\frac{\hat{\kappa}_{T}\(\hat{\mu}_{1}^2 + \hat{\mu}_{2}^2\)}{\hat{\eta}\hat{T}} = \(\frac{2\pi}{\s m}\)^2\,.
\ee
Again, using the results in \cite{ouroldpaper} for the single charge case, we obtain the same value for this ratio (with $\hat{\mu}_{1}^2 + \hat{\mu}_{2}^2$ replaced by $\hat{\mu}^2$).

\subsubsection{Discussion of various bounds}

Note that the transport coefficients (\ref{ShearViscosity})-(\ref{BulkViscosity}) are universally valid for any solution with the same asymptotics as the black brane solution discussed in the previous section. 

We first consider the bulk to shear viscosity ratio, given by (\ref{ratio_1}), whose value is fixed kinematically by the reduction, and will be different depending on the asymptotics of the system. This is also true for the ratio $\hat{\k}_T/\hat{\eta}$.

As in \cite{ouroldpaper}, the bound proposed in \cite{Buchel:2007mf} for the bulk to shear viscosity ratio 
\be \label{bound}
    \frac{\hat\zeta_s}{\hat\eta} \geq  2\left(\frac1{d-1}-\hat c_{s}^2\right)\,,
\ee
doesn't hold. And again, for a general system there is no reason to expect that such a bound would necessarily apply, since $\hat\zeta_s/\hat\eta$ is fixed kinematically. Indeed, rewriting (\ref{ratio_1}) using (\ref{sos}), we obtain
\be
\frac{\hat\zeta_s}{\hat\eta} = 2\left(\frac1{d-1}-\hat c_{s}^2\right) -\frac{4\((\s-1)\cosh^4\w_1\cosh^4\w_2 +(2-\s)\cosh^2\w_1\cosh^2\w_2 -1\) }{\(2(\s-1)\cosh^2\w_1\cosh^2\w_2 +1\)^2}\,,
\ee
so that clearly this bound is always violated, except if
\be
    \s<\hat{\mu}_{1}^2 + \hat{\mu}_{2}^2\,.
\ee
Since $\hat{\mu}_{1}^2 + \hat{\mu}_{2}^2 = 1-1/(\cosh^2\w_1\cosh^2\w_2) \leq 1$, the above is only possible if $\s <1$. The equality is achieved when either $\hat{\mu}_{1}=\hat{\mu}_{2}=0$ (neutral case) or else $\hat{\mu}_{1}^2 + \hat{\mu}_{2}^2=\sigma$.

However, an inequality which is satisfied in this case, and was also satisfied in \cite{ouroldpaper}, is
\be \label{bound2}
    \frac{\hat\zeta_s}{\hat\eta} \geq  2\left(\frac1{d-1}-\hat c_{q}^2\right)\,,
\ee
where for our case
\be
\hat c_q^2 \equiv \left.\frac{\partial \hat P}{\partial \hat\epsilon}\right|_{\hat q_1,\hat q_2}  = \frac{2\cosh^2\w_1\cosh^2\w_2 -1}{2(\s-1)\cosh^2\w_1\cosh^2\w_2 +1}\,.
\ee
$\hat c_q^2$ reduces to the speed of sound of the conformal branes when $\w_1=\w_2=0$. If we rewrite our ratio as follows
\be
\frac{\hat\zeta_s}{\hat\eta} - 2\left(\frac1{d-1} -\hat c_{q}^2\right) = \frac{4(\s-1)\cosh^2\w_1\cosh^2\w_2(\cosh^2\w_1\cosh^2\w_2 -1)}{\(2(\s-1)\cosh^2\w_1\cosh^2\w_2+1\)^2}\,,
\ee
we see that the right hand side is manifestly positive when $\s > 1$. It still remains interesting to check whether there are any counterexamples to this inequality.

We also note that the KSS bound is saturated for this dual charge system, since we may use (\ref{ShearViscosity}) and (\ref{shat}) to see that
\be
\frac{\hat{\eta}}{\hat{s}}=\frac{1}{4\pi}\,,
\ee
which results due to the fact that this bound is saturated for conformal branes. The same holds true for the uncharged case, and results due to the fact that in all these cases we require regularity in the interior.

\section{Discussion and conclusions}

In this paper, we use generalized dimensional reduction of AdS gravity to determine the holographic dictionary of a specific EMD theory containing two gauge fields, three neutral scalars and an axion. We also study the hydrodynamic behaviour of the dual theory, and compute its first order transport coefficient. Such an analysis was performed in \cite{ouroldpaper} for a reduced EMD theory with one gauge field and two scalars. We reproduce all results in \cite{ouroldpaper} by removing the extra fields from the results in this paper.

In contrast to \cite{ouroldpaper}, when considering the universal hydrodynamics of the reduced theory, we find that the presence of the extra charge in the system leads us, \emph{via} the modified frame-independent analysis of \cite{Bhattacharya:2011tr}, to a matrix of conductivities, from which we may then calculate the thermal conductivity. In this paper we also have to modify the formula of \cite{Eling:2011ms} used to check the bulk to shear viscosity ratio, since that formula applies to the case of canonically normalized  scalars and we have a non-standard axion normalization. Lastly, the system studied in this paper also satisfies the the modified bound on the bulk to shear viscosity ratio found in \cite{ouroldpaper}.

There are numerous possible extensions to this work, the most obvious being the generalization of this analysis to include many gauge fields, which may be useful in the study of imbalanced superconductors. Making connections to similar systems in the context of AdS/CMT or cosmology are other possible avenues of future work.

\section*{Acknowledgements}

I would like to thank Kostas Skenderis, Marika Taylor and Blaise Gout\'eraux for discussions. This work is part of a research program which is financially supported by the `Nederlandse Organisatie voor Wetenschappelijk Onderzoek' (NWO). MS acknowledges support via the NWO Vici grant of Kostas Skenderis.

\appendix
\section {Appendix}

\subsection{Equations of motion of reduced theory with two Maxwell fields }\label{app:eoms}

In this appendix we give the equations of motion for each of the fields stemming from the action (\ref{redactfull}).

The equation of motion for $\z$ is
\bea
\nabla_M [e^{\y} \pa^M \z] &=& \frac{1}{4} (2 \s  -d -1)e^{\y}  \( e^{\frac{2 (\y + \z)}{(2 \s-d)}} F_{MN}^{(2)} F^{(2)MN}\right.\nn\\
&&\left. -\frac{1}{2\s-d-1}e^{\frac{2(\x(2\s-d)+(2\s-d-1)\y-\z)}{(2\s-d)(2\s-d-1)}}\times \right. \nn\\
&&\left. \times\(F_{MN}^{(1)}F^{(1)MN} + 4\pa^M A^{(0)}A^{(2)N}F_{MN}^{(1)} -2\(A^{(2)M}\pa_M A^{(0)}\)^{2} \right.\right.  \nn\\
&&\qquad\qquad \left.\left. +2\(e^{\frac{-2 (\y + \z)}{(2 \s-d)}}+A_{M}^{(2)}A^{(2)M}\)\(\pa A^{(0)}\)^{2} \)\right. \nn\\
&& \left. -2e^{\frac{2(\x-\z)}{(2\s-d-1)}}\pa_N A^{(0)}\pa^N A^{(0)} \)\,,
\eea
while the equation of motion for $\y$ is
\bea
\nabla_M [e^{\y} \pa^M \y] &=& \frac{2\s-d}{2(2 \s  -d -1)}e^{\y}  \left[ R_{(d+1)} + \frac{(2\s-d-1)}{(2 \s -d)} (\partial \y)^2 \right. \nn\\
&& \left. - \frac{1}{(2 \s -d)(2\s-d-1)} (\partial \z)^2 - \frac{1}{(2 \s -d-1)(2\s-d-2)} (\partial \x)^2 \right.  \nn\\
&&\left. + 2\s(2\s-1)  -\frac{1}{4}\frac{(2\s-d+2)}{(2\s-d)} e^{\frac{2 (\y + \z)}{(2 \s-d)}} F_{MN}^{(2)} F^{(2)MN}\right.\nn\\
&&\left. -\frac{1}{4}\frac{2\s-d+2}{2\s-d}e^{\frac{2(\x(2\s-d)+(2\s-d-1)\y-\z)}{(2\s-d)(2\s-d-1)}}\times \right. \nn\\
&&\left. \times\(F_{MN}^{(1)}F^{(1)MN} + 4\pa^M A^{(0)}A^{(2)N}F_{MN}^{(1)} -2\(A^{(2)M}\pa_M A^{(0)}\)^{2} \right.\right. \nn\\
&&\qquad\qquad \left.\left. +2\(e^{\frac{-2 (\y + \z)}{(2 \s-d)}}+A_{M}^{(2)}A^{(2)M}\)\(\pa A^{(0)}\)^{2} \)\right.\nn\\
&&\left. +\frac{1}{2\s-d}e^{\frac{2(\x-\z)}{(2\s-d-1)}}\pa_N A^{(0)}\pa^N A^{(0)} \right]\,,
\eea
and finally for $\x$ 
\bea
\nabla_M [e^{\y} \pa^M \x] &=&-\frac{1}{4}(2\s-d-2)e^{\y}e^{\frac{2(\x(2\s-d)+(2\s-d-1)\y-\z)}{(2\s-d)(2\s-d-1)}}\times \nn\\
&&\times\(F_{MN}^{(1)}F^{(1)MN} + 4\pa^M A^{(0)}A^{(2)N}F_{MN}^{(1)} -2\(A^{(2)M}\pa_M A^{(0)}\)^{2} \right. \nn\\
&&\qquad\qquad \left. +2\(e^{\frac{-2 (\y + \z)}{(2 \s-d)}}+A_{M}^{(2)}A^{(2)M}\)\(\pa A^{(0)}\)^{2} \)\,.
\eea

The gravitational field equation is
\bea
0&=&R_{MN} -\half g_{MN}R_{(d+1)} - \nabla_N\pa_M\y +\nabla^P\pa_P\y\ g_{MN} \nn\\
&&-\half g_{MN}\( \frac{-(2\s-d+1)}{(2 \s -d)} (\partial \y)^2 - \frac{1}{(2 \s -d)(2\s-d-1)} (\partial \z)^2 \right. \nn\\
&&\left.  \qquad\qquad- \frac{1}{(2 \s -d-1)(2\s-d-2)} (\partial \x)^2 -\frac{1}{4} e^{\frac{2 (\y + \z)}{(2 \s-d)}} F_{PQ}^{(2)} F^{(2)PQ}\right.  \nn\\
&&\left. \qquad\qquad +2\s(2\s-1)-\frac{1}{4}e^{\y}e^{\frac{2(\x(2\s-d)+(2\s-d-1)\y-\z)}{(2\s-d)(2\s-d-1)}}\times \right.\nn\\
&&\left. \qquad\qquad\times\(F_{PQ}^{(1)}F^{(1)PQ} + 4\pa^P A^{(0)}A^{(2)Q}F_{PQ}^{(1)} -2\(A^{(2)P}\pa_P A^{(0)}\)^{2} \right.\right. \nn\\
&&\qquad\qquad \qquad\left. \left. +2\(e^{\frac{-2 (\y + \z)}{(2 \s-d)}}+A_{P}^{(2)}A^{(2)P}\)\(\pa A^{(0)}\)^{2} \)\) \nn\\
&& -\frac{1}{2\s-d}\pa_M \y\pa_N\y -\frac{1}{(2\s-d)(2\s-d-1)}\pa_M \z\pa_N\z \nn\\
&&-\frac{1}{(2\s-d-1)(2\s-d-2)}\pa_M \x\pa_N\x - \half e^{\frac{2 (\y + \z)}{(2 \s-d)}} F_{M}^{(2)Q} F_{NQ}^{(2)} \nn\\
&&-\frac{1}{4}e^{\frac{2(\x(2\s-d)+(2\s-d-1)\y-\z)}{(2\s-d)(2\s-d-1)}}\times \nn\\
&&\times \( 2F_{MP}^{(1)}F_{N}^{(1)P} +4\( \pa_N A^{(0)}A^{(2)P}-\pa^P A^{(0)}A_N^{(2)}\)F_{MP}^{(1)} \right. \nn\\
&& \left.\qquad-4A_{N}^{(2)}A^{(2)P}\pa_P A^{(0)}\pa_M A^{(0)} +2e^{\frac{-2(\y+\z)}{(2\s-d)}}\pa_M A^{(0)}\pa_N A^{(0)} \right. \nn\\
&& \left.\qquad+2\( A_M^{(2)}A_N^{(2)}\pa_P A^{(0)}\pa^P A^{(0)} + A_P^{(2)}A^{(2)P}\pa_M A^{(0)} \pa_N A^{(0)} \) \)\,, \nn\\
&&\mbox{\ \ }
\eea
the gauge field equations are
\bea
&&\nabla_M\left[e^{\y}e^{\frac{2(\x(2\s-d)+(2\s-d-1)\y-\z)}{(2\s-d)(2\s-d-1)}}\( F^{(1)MN}+\pa^M A^{(0)}A^{(2)N}-\pa^N A^{(0)} A^{(2)M}\)\right] = 0\,, \nn\\
&&\mbox{\ \ }
\eea
and
\bea
\nabla_M\left[e^{\y}e^{\frac{2(\y+\z)}{(2\s-d)}}F^{(2)MN}\right] &=& e^{\y}e^{\frac{2(\x(2\s-d)+(2\s-d-1)\y-\z)}{(2\s-d)(2\s-d-1)}} \( \pa^M A^{(0)} F_M^{(1)N}\right. \nn\\
&&\left.  - A_M^{(2)}\pa^M A^{(0)}\pa^N A^{(0)} + A^{(2)N}\(\pa A^{(0)}\)^{2} \)\,, \nn\\
&&\mbox{\ \ }
\eea
and the axion field equation is
\bea
&&\nabla_M\left[ e^{\y}e^{\frac{2(\x(2\s-d)+(2\s-d-1)\y-\z)}{(2\s-d)(2\s-d-1)}} \( F_{\ \ \ \ \ N}^{(1)M} A^{(2)N} - A^{(2)M}A^{(2)N}\pa_N A^{(0)}  \right. \right. \nn\\
&& \left. \left. \qquad\qquad+ \( e^{\frac{-2(\y+\z)}{(2\s-d)}}+A_N^{(2)}A^{(2)N}\) \pa^M  A^{(0)} \) \right] =0\,.
\eea
Poles exist in the equations of motion for $2\s=d$, so we have to go back to the reduction ansatz to see that this corresponds to the case where there is no reduction. Furthermore, for $2\s=d+1$ the reduction is along a circle and there are no additional scalar fields $\z$ and $\xi$, while $2\s=d+2$ corresponds to a reduction along $\mathbf T^{2}$ with no scalar field $\x$.

\subsection{Check of quantities sourced by non-normalizable modes of fields} \label{app:inverse_zero_metric}

In this appendix we will write down the explicit component forms of $g_{(0)\mu\nu}$ and $g_{(0)}^{\mu\nu}$, and then we will show that $A_{(0)}^{(3)i}$ sources current $\< \hat{J}^{(1)i} \>_d$, while $\< \hat{J}^{(2)i} \>_d$ is the current sourced by $A_{(0)}^{(2)i}$. Showing that the scalars $\(\k_{(0)}, \z_{(0)}, \x_{(0)}, A_{(0)}^{(0)}\)$ respectively source $\(\cO_{\y}, \cO_{\zeta}, \cO_{\x}, \cO_{A^{(0)}}\)$ follows along the same lines.

We first write down the elements of $g_{(0)\mu\nu}$ in (\ref{AdS_exp}), using the anzats (\ref{redansfull}), as well as expansions (\ref{AdS_exp_metric}) - (\ref{AdS_exp_axion}). More precisely, we need to find the coefficient of $\r^{0}$ in (\ref{redansfull}).

This gives
\bea
g_{(0)ij} &=& \tilde{g}_{(0)ij}+e^{\Lambda(\k_{(0)}+\z_{(0)})}A_{i(0)}^{(2)}A_{j(0)}^{(2)} + e^{\Lambda\k_{(0)}+2\Theta\x_{(0)}-\Lambda\Theta\z_{(0)}}A_{i(0)}^{(1)}A_{j(0)}^{(1)}\,,\nn\\
g_{(0)iy_1} &=& -e^{\Lambda\k_{(0)}+2\Theta\x_{(0)}-\Lambda\Theta\z_{(0)}}A_{i(0)}^{(1)} \,,\nn\\
g_{(0)iy_2} &=& -e^{\Lambda(\k_{(0)}+\z_{(0)})}A_{i(0)}^{(2)}+e^{\Lambda\k_{(0)}+2\Theta\x_{(0)}-\Lambda\Theta\z_{(0)}}A_{i(0)}^{(1)}A_{(0)}^{(0)}\,,\nn\\
g_{(0)y_1y_1} &=&  e^{\Lambda\k_{(0)}+2\Theta\x_{(0)}-\Lambda\Theta\z_{(0)}}\,,\nn\\ 
g_{(0)y_2y_2} &=& e^{\Lambda(\k_{(0)}+\z_{(0)})} + e^{\Lambda\k_{(0)}+2\Theta\x_{(0)}-\Lambda\Theta\z_{(0)}}A_{(0)}^{(0)2}\,,\nn\\
g_{(0)y_1 y_2} &=& -e^{\Lambda\k_{(0)}+2\Theta\x_{(0)}-\Lambda\Theta\z_{(0)}}A_{(0)}^{(0)}\,, \nn\\
g_{(0)ab}&=& e^{\Lambda\k_{(0)}-\Lambda\Theta\z_{(0)}-2\Theta\Omega\x_{(0)}}\d_{ab}\,, \label{boundmetric}
\eea
where $\Lambda = \frac{2}{(2\s-d)}$ and $\Theta = \frac{1}{(2\s-d-1)}$, while $\Omega = \frac{1}{(2\s-d-2)}$.

With these values in hand, it is now easy to check that $g^{\mu\nu}_{(0)}$, the inverse metric of (\ref{boundmetric}), is given by 
\bea\label{metricinverse}
&& g_{(0)}^{ij} = \tilde{g}_{(0)}^{ij}\,, \qquad g_{(0)}^{iy^{(1)}}=A_{(0)}^{(3)i}\,, \qquad g_{(0)}^{iy^{(2)}} = A_{(0)}^{(2)i}\,, \nn\\
&& g_{(0)}^{y_{1}y_{1}} = e^{-\frac{2\((2\s-d-1)\k\sub{0}+(2\s-d)\x\sub{0}-\z\sub{0}\)}{(2\s-d)(2\s-d-1)}} + e^{-\frac{2\(\k\sub{0} + \z\sub{0}\)}{2 \s-d}} A_{(0)}^{(0)\ 2} + A_{(0)n}^{(3)}A_{(0)}^{(3)n}\,, \nn\\
&& g_{(0)}^{y_{1}y_{2}} = e^{-\frac{2\(\k\sub{0} + \z\sub{0}\)}{2 \s-d}} A_{(0)}^{(0)} + A_{(0)n}^{(3)}A_{(0)}^{(2)n}\,, \qquad g_{(0)}^{y_{2}y_{2}} = e^{-\frac{2\(\k\sub{0} + \z\sub{0}\)}{2 \s-d}}  + A_{(0)n}^{(2)}A_{(0)}^{(2)n}\,, \nn\\
&& g_{(0)}^{ab} = e^{-\frac{2\((2\s-d-1)(2\s-d-2)\k\sub{0}-(2\s-d-2)\z\sub{0}-(2\s-d)\x\sub{0}\)}{(2\s-d)(2\s-d-1)(2\s-d-2)}}\d^{ab}\,, \nn\\
\eea
and indeed satisfies $g_{(0)\mu\lambda}g_{(0)}^{\lambda\nu} = \d_{\mu}^{\nu}$. We may write this inverse metric (\ref{metricinverse}) more simply by setting
\bea\label{alphabetagamma}
&&\a_{(0)} = \Lambda(\k_{(0)} + \z_{(0)})\,,\qquad \b_{(0)} = \Lambda\k_{(0)} +2\Theta\x_{(0)}-\Lambda\Theta\z_{(0)}\,, \nn\\
&&\qquad\qquad\qquad\gamma_{(0)} = \Lambda\k_{(0)} -\Lambda\Theta\z_{(0)} - 2\Theta\x_{(0)}\,,\nn\\
\eea
so that
\bea
&& g_{(0)}^{ij} = \tilde{g}_{(0)}^{ij}\,, \qquad g_{(0)}^{iy^{(1)}}=A_{(0)}^{(3)i}\,, \qquad g_{(0)}^{iy^{(2)}} = A_{(0)}^{(2)i}\,, \nn\\
&& g_{(0)}^{y_{1}y_{1}} = e^{-\b_{(0)}} + e^{-\a_{(0)}} A_{(0)}^{(0)\ 2} + A_{(0)n}^{(3)}A_{(0)}^{(3)n}\,, \nn\\
&& g_{(0)}^{y_{1}y_{2}} = e^{-\a_{(0)}} A_{(0)}^{(0)} + A_{(0)n}^{(3)}A_{(0)}^{(2)n}\,, \qquad g_{(0)}^{y_{2}y_{2}} = e^{-\a_{(0)}}  + A_{(0)n}^{(2)}A_{(0)}^{(2)n}\,, \nn\\
&& g_{(0)}^{ab} = e^{-\gamma_{(0)}}\d^{ab}\,. \nn\\
\eea
We also set
\bea
&&\a_{(2\s)} = \Lambda(\k_{(2\s)} + \z_{(2\s)})\,,\qquad \b_{(2\s)} = \Lambda\k_{(2\s)} +2\Theta\x_{(2\s)}-\Lambda\Theta\z_{(2\s)}\,, \nn\\
&&\qquad\qquad\qquad\gamma_{(2\s)} = \Lambda\k_{(2\s)} -\Lambda\Theta\z_{(2\s)} - 2\Theta\x_{(2\s)}\,.\nn
\eea
We may now move on to the quantities sourced by $A_{(0)}^{(2)i}$ and $A_{(0)}^{(3)i}$. Now, recall that
\be
\<T_{\mu\nu}\>_{2\s} = \frac{2}{\sqrt{g_{(0),2\s}}}\frac{\d S_{ren}}{\d g_{(0)}^{\mu\nu}}\,, \qquad \<t_{\mu\nu}\>_{d} = e^{\k_{(0)}}(2\pi R)^{2\s-d}\<T_{\mu\nu}\>_{2\s}\,,
\ee
so that the current ${\cal J}^{(2)}_i$ sourced by $A_{(0)}^{(2)i}$ is given by
\bea
\<{\cal J}^{(2)}_i\>_d &\equiv& \frac{1}{\sqrt{g_{(0),d}}}\frac{\d S_{ren}}{\d A_{(0)}^{(2)i}} = \frac{e^{\k_{(0)}}}{\sqrt{g_{(0),2\s}}}\frac{\d S_{ren}}{\d g_{(0)}^{\k\r}}\frac{\d g_{(0)}^{\k\r}}{\d A_{(0)}^{(2)i}} = \frac{1}{2}e^{\k_{(0)}}\<T_{\k\r}\>_{2\s}\frac{\d g_{(0)}^{\k\r}}{\d A_{(0)}^{(2)i}} \nn\\
&=& \frac{1}{2}(2\pi R)^{d-2\s}\<t_{\k\r}\>_d \frac{\d g_{(0)}^{\k\r}}{\d A_{(0)}^{(2)i}}\nn\\
&=& \frac{1}{2}(2\pi R)^{d-2\s} \(2\<t_{jy_{2}}\>_d\frac{\d g_{(0)}^{jy_{2}}}{\d A_{(0)}^{(2)i}} + 2 \<t_{y_{1}y_{2}}\>_d\frac{\d g_{(0)}^{y_{1}y_{2}}}{\d A_{(0)}^{(2)i}} +\<t_{y_{2}y_{2}}\>_d\frac{\d g_{(0)}^{y_{2}y_{2}}}{\d A_{(0)}^{(2)i}} \) \nn \\
&=& \frac{1}{2}(2\pi R)^{d-2\s} \(2\<t_{iy_{2}}\>_d + 2\<t_{y_{1}y_{2}}\>_d A_{(0)i}^{(3)}+ 2\<t_{y_{2}y_{2}}\>_dA_{(0)i}^{(2)}\) \nn\\
&=& (2\pi R)^{d-2\s}2\s L e^{\k_{(0)}} \nn\\
&& \times\left[ e^{\b_{(0)}}\(\b_{(2\s)}A_{i(0)}^{(1)}A_{(0)}^{(0)} + A_{i(2\s)}^{(1)}A_{(0)}^{(0)} + A_{(2\s)}^{(0)}A_{i(0)}^{(1)}\)\right.\nn\\
&& \left.\qquad-e^{\a_{(0)}}\(\a_{(2\s)} A_{i(0)}^{(2)}+A_{i(2\s)}^{(2)}\) -A_{i(0)}^{(3)}e^{\b_{(0)}}\(\b_{(2\s)}A_{(0)}^{(0)} + A_{(2\s)}^{(0)}\) \right. \nn\\
&&\left. \qquad+ A_{i(0)}^{(2)}\(e^{\a_{(0)}}\a_{(2\s)}+e^{\b_{(0)}}\(A_{(0)}^{(0)2}\b_{(2\s)}+2A_{(0)}^{(0)}A_{(2\s)}^{(0)}\)\) \right] \nn\\
&=& (2\pi R)^{d-2\s}2\s L e^{\k_{(0)}}\left[ -e^{\a_{(0)}}A_{i(2\s)}^{(2)} + e^{\b_{(0)}}\(A_{i(2\s)}^{(1)}A_{(0)}^{(0)} + A_{(2\s)}^{(0)}A_{i(0)}^{(1)} \right.\right. \nn\\
&&\left. \left. \qquad\qquad\qquad\qquad\qquad- A_{i(0)}^{(3)}A_{(2\s)}^{(0)} + 2A_{(0)}^{(0)}A_{i(0)}^{(2)}A_{(2\s)}^{(0)}\)\right] \nn\\
&=& -(2\pi R)^{d-2\s}2\s L e^{\k_{(0)}}\left[ e^{\a_{(0)}}A_{i(2\s)}^{(2)} - e^{\b_{(0)}}A_{(0)}^{(0)}\(A_{i(2\s)}^{(1)}+A_{i(0)}^{(2)}A_{(2\s)}^{(0)}\)\right]\nn\\
&=& \<\hat{J}^{(2)}_i\>_d\,,
\eea
while the current ${\cal J}^{(1)}_i$ sourced by $A_{(0)}^{(3)i}$ is 
\bea
\<{\cal J}^{(1)}_i\>_d &\equiv&\frac{1}{\sqrt{g_{(0),d}}}\frac{\d S_{ren}}{\d A_{(0)}^{(3)i}}=\frac{1}{\sqrt{g_{(0),d}}}\frac{\d S_{ren}}{\d g_{(0)}^{\k\r}}\frac{\d g_{(0)}^{\k\r}}{\d A_{(0)}^{(3)i}}=\frac{1}{2}(2\pi R)^{d-2\s}\<t_{\k\r}\>_d \frac{\d g_{(0)}^{\k\r}}{\d A_{(0)}^{(3)i}}\nn\\
&&\qquad \qquad= \frac{1}{2}(2\pi R)^{d-2\s} \(2\<t_{jy_{1}}\>_d\frac{\d g_{(0)}^{jy_{1}}}{\d A_{(0)}^{(3)i}} + \<t_{y_{1}y_{1}}\>_d\frac{\d g_{(0)}^{y_{1}y_{1}}}{\d A_{(0)}^{(3)i}} + 2 \<t_{y_{1}y_{2}}\>_d\frac{\d g_{(0)}^{y_{1}y_{2}}}{\d A_{(0)}^{(3)i}} \) \nn \\
&&\qquad\qquad = \frac{1}{2}(2\pi R)^{d-2\s} \(2\<t_{iy_{1}}\>_d + 2\<t_{y_{1}y_{1}}\>_d A_{(0)i}^{(3)}+ 2\<t_{y_{1}y_{2}}\>_dA_{(0)i}^{(2)}\) \nn\\
&& \qquad\qquad = -(2\pi R)^{d-2\s}2 \s L e^{\k\sub{0}} e^{\frac{2\((2\s-d-1)\k\sub{0}+(2\s-d)\x\sub{0}-\z\sub{0}\)}{(2\s-d)(2\s-d-1)}} \( A_{i(2\s)}^{(1)} + A_{(2\s)}^{(0)}A_{i(0)}^{(2)}  \) \nn\\
&&\qquad\qquad = \< \hat{J}^{(1)}_i \>_d\,.\label{Sren_check}
\eea

\subsection{Transport coefficient relations in two charge hydrodynamic system}\label{app:transcoeffs2}

In this appendix we derive the equations from which we may extract the transport coefficients for a two-charge hydrodynamic system, using the frame-independent method of \cite{Bhattacharya:2011tr}. Note that we may reproduce the one-charge results of \cite{Bhattacharya:2011tr} by setting $\mu_2 = F_{\mu\nu}^{(2)} =0$, while $\mu_1=\mu$ and $F_{\mu\nu}^{(1)}\to -F_{\mu\nu}$.

In \cite{Bhattacharya:2011tr}, beginning with a one charge system in 4 dimensions with degrees of freedom $\(\mu, T, u_{\mu}\)$ and entropy current divergence
\be\label{BhatJs}
\nabla_{\mu}J^{\mu}_s = -\nabla_{\mu}\(\frac{u_{\nu}}{T}\)T_{\rm diss}^{\mu\nu} - \(\partial_{\mu}\frac{\mu}{T} - \frac{F_{\mu\nu}u^{\nu}}{T}\)J_{\rm diss}^{\mu}\,,
\ee
the first step involves writing down all possible scalars, vectors and tensors in the theory, and then expanding each term on the RHS of (\ref{BhatJs}) in terms of these. Having found which quantities participate in this expansion, the next step involves using the first order equations of motion of the system to show that the scalars and vectors are proportional to each other, respectively.

The procedure we employ is equivalent, except that we extend the analysis to include an extra chemical potential (as well as an extra field strength and current). Furthermore, we will be using the notation developed in the rest of this paper, namely hatted quantities and Latin indices, as opposed to the Greek indices used in \cite{Bhattacharya:2011tr}.

Thus, our $d$-dimensional two charge system has degrees of freedom $(\hat{\mu}_1, \hat{\mu}_2, \hat{T}, \hat{u}_{i})$ and entropy current divergence given by
\bea\label{app3_1}
\nabla_{i}J_{s}^{i} &=&  -\nabla_{i}\(\frac{\hat{u}_{j}}{\hat{T}}\)\hat{T}_{\rm diss}^{ij} - \(\partial_{i}\(\frac{\hat{\mu}_1}{\hat{T}}\)-\frac{F^{(1)}_{(0)ji}\hat{u}^{j}}{\hat{T}}\)\hat{J}_{\rm diss}^{(1)i} \nn\\
&&\qquad\qquad\qquad- \(\partial_{i}\(\frac{\hat{\mu}_2}{\hat{T}}\)-\frac{F^{(2)}_{(0)ji}\hat{u}^{j}}{\hat{T}}\)\hat{J}_{\rm diss}^{(2)i} \,,
\eea
with all possible scalars, vectors and tensors in the theory given by the obvious extension of Table 1 in \cite{Bhattacharya:2011tr}. Expanding the RHS of (\ref{app3_1}) in terms of these, we get
\bea
\nabla_{i}J_{s}^{i} &=& \hat{T}_{\rm diss}^{ij}\left[ \frac{-\hat{P}_{ij}}{(d-1)}\frac{(\partial\cdot \hat{u})}{\hat{T}} - \frac{(\hat{u}\cdot\partial)\hat{T}}{\hat{T}^2}\hat{u}_{i}\hat{u}_{j}+\frac{1}{\hat{T}}\(\hat{P}_{ik}\frac{\partial^{k}\hat{T}}{\hat{T}}+(\hat{u}\cdot\partial)\hat{u}_{i}\)\hat{u}_{j} - \frac{\hat{\s}_{ij}}{\hat{T}}\right] \nn\\
&& + \hat{J}_{\rm diss}^{(1)i} \left[ (\hat{u}\cdot\partial)\hat{\nu}_1 \hat{u}_{i}+ V_{1i}^{(1)}\right] + \hat{J}_{\rm diss}^{(2)i} \left[ (\hat{u}\cdot\partial)\hat{\nu}_2 \hat{u}_{i}+ V_{1i}^{(2)}\right]\,, \label{app3_3}
\eea
where we have defined $\hat{\nu}_1\equiv \hat{\mu}_1/\hat{T}$ and $\hat{\nu}_2\equiv \hat{\mu}_2/\hat{T}$ for simplicity, and $V_{1i}^{I}$ $(I=1,2)$ is as defined in (\ref{Veqn}). It is clear from this that the $SO(d-1)$-invariant quantities involved in the expansion are 
\bea
&&\mbox{scalars:}\qquad \frac{(\hat{u}\cdot\partial)\hat{T}}{\hat{T}}\,,\qquad (\hat{u}\cdot\partial)\hat{\nu}_1\,,\qquad (\hat{u}\cdot\partial)\hat{\nu}_2\,, \qquad (\partial\cdot \hat{u})\,, \nn\\
&&\mbox{vectors:}\qquad V_{1i}^{(1)}\,, \qquad V_{1i}^{(2)}\,,\qquad V_{1i}^{(3)}\equiv\(\hat{P}_{ik}\frac{\partial^{k}\hat{T}}{\hat{T}}+(\hat{u}\cdot\partial)\hat{u}_{i}\)\,,\nn\\
&&\mbox{tensor:}\qquad \hat{\s}_{ij}\,.\label{app3_4}
\eea 

Now, to see that all the scalars may be written in terms of $(\partial\cdot \hat{u})$, while the vector $V_{1}^{(3)}$  may be written as a linear combination of $V_{1}^{(1)}$ and $V_{1}^{(2)}$, we also use the equations of motion
\be\label{app3_5}
\partial_{i}\hat{T}^{ij} = F^{(1)ij}_{(0)}\hat{J}_{i}^{(1)} +  F^{(2)ij}_{(0)}\hat{J}_{i}^{(2)}\,,\qquad\partial_{i}\hat{J}^{(1)i} = 0\,,\qquad\partial_{i}\hat{J}^{(2)i} = 0\,, 
\ee
with
\be\label{app3_6}
\hat{T}^{ij} = (\hat{\e} +\hat{P})\hat{u}^{i}\hat{u}^{j} +\hat{P}\hat{\eta}^{ij} + \hat{T}^{ij}_{\rm diss}\,,\qquad \hat{J}^{(1)i} = \hat{q}_1 u^{i} + \hat{J}_{\rm diss}^{(1)i}\,, \qquad \hat{J}^{(2)i} = \hat{q}_2 u^{i} + \hat{J}_{\rm diss}^{(2)i}\,, 
\ee
so that at first order
\bea
\hat{u}_{j}\partial_{i}\hat{T}^{ij} = 0 &\Rightarrow& (\hat{u}\cdot\partial)\hat{\e} = - (\hat{\e}+\hat{P})(\partial\cdot \hat{u}) \,,\label{app3_7}\\
\partial_{i}\hat{J}^{(1)i} = 0 &\Rightarrow& (\hat{u}\cdot\partial)\hat{q}_1 = -\hat{q}_1 (\partial\cdot \hat{u})\,,\label{app3_8}\\
\partial_{i}\hat{J}^{(2)i} = 0 &\Rightarrow& (\hat{u}\cdot\partial)\hat{q}_2 = -\hat{q}_2 (\partial\cdot \hat{u})\,.\label{app3_9}
\eea

By making various combinations of the equations (\ref{app3_7}) - (\ref{app3_9}) above, following the example of \cite{Bhattacharya:2011tr}, and simplifying the RHS and LHS of these resulting equations, we may show that 
\bea
\frac{(\hat{u}\cdot\partial)\hat{T}}{\hat{T}} &=& -\left . \(\frac{\partial \hat{P}}{\partial \hat{\e}}\)\right |_{\hat{q}_1,\hat{q}_2}(\partial\cdot \hat{u})\,,\label{app3_20} \\
(\hat{u}\cdot\partial)\hat{\nu}_1 &=& -\frac{1}{\hat{T}}\left. \(\frac{\partial \hat{P}}{\partial \hat{q}_1}\) \right|_{\hat{\e},\hat{q}_2}(\partial\cdot \hat{u})\,,\label{app3_25}\\
(\hat{u}\cdot\partial)\hat{\nu}_2 &=& -\frac{1}{\hat{T}}\left. \(\frac{\partial \hat{P}}{\partial \hat{q}_2}\) \right|_{\hat{\e},\hat{q}_1}(\partial\cdot \hat{u})\,.\label{app3_28}
\eea
To obtain (\ref{app3_20}), we take the combination
\bea
\left . \(\frac{\partial \hat{P}}{\partial \hat{\e}}\)\right |_{\hat{q}_1,\hat{q}_2}&&\left\{\left . \(\frac{\partial \hat{q}_2}{\partial\hat{\nu}_2}\)\right|_{\hat{T},\hat{\nu}_1}\left[\left . \(\frac{\partial \hat{q}_1}{\partial\hat{\nu}_1}\)\right|_{\hat{T},\hat{\nu}_2}\times(\ref{app3_7}) - \left . \(\frac{\partial\hat{\e}}{\partial\hat{\nu}_1}\)\right|_{\hat{T},\hat{\nu}_2}\times (\ref{app3_8})\right] \right.\nn\\
&+& \left.\left . \(\frac{\partial\hat{\e}}{\partial\hat{\nu}_2}\)\right|_{\hat{T},\hat{\nu}_1}\left[\left . \(\frac{\partial \hat{q}_2}{\partial\hat{\nu}_1}\)\right|_{\hat{T},\hat{\nu}_2}\times(\ref{app3_8}) - \left . \(\frac{\partial \hat{q}_1}{\partial\hat{\nu}_1}\)\right|_{\hat{T},\hat{\nu}_2}\times (\ref{app3_9})\right]\right. \nn\\
&-&\left.\left . \(\frac{\partial \hat{q}_1}{\partial\hat{\nu}_2}\)\right|_{\hat{T},\hat{\nu}_1}\left[\left . \(\frac{\partial \hat{q}_2}{\partial\hat{\nu}_1}\)\right|_{\hat{T},\hat{\nu}_2}\times(\ref{app3_7}) - \left . \(\frac{\partial\hat{\e}}{\partial\hat{\nu}_1}\)\right|_{\hat{T},\hat{\nu}_2}\times (\ref{app3_9})\right]\right\}\,,\label{app3_10}\nn\\
\eea
while for (\ref{app3_25}) we use
\bea
\left. \(\frac{\partial \hat{P}}{\partial \hat{q}_1}\)\right|_{\hat{\e},\hat{q}_2}&&\left\{\left . \(\frac{\partial \hat{q}_2}{\partial\hat{\nu}_2}\)\right|_{\hat{T},\hat{\nu}_1}\left[\left . \(\frac{\partial \hat{q}_1}{\partial \hat{T}}\)\right|_{\hat{\nu}_1,\hat{\nu}_2}\times(\ref{app3_7}) - \left . \(\frac{\partial\hat{\e}}{\partial \hat{T}}\)\right|_{\hat{\nu}_1,\hat{\nu}_2}\times (\ref{app3_8})\right]\right. \nn\\
&+&\left. \left . \(\frac{\partial\hat{\e}}{\partial\hat{\nu}_2}\)\right|_{\hat{T},\hat{\nu}_1}\left[\left . \(\frac{\partial \hat{q}_2}{\partial \hat{T}}\)\right|_{\hat{\nu}_1,\hat{\nu}_2}\times(\ref{app3_8}) - \left . \(\frac{\partial \hat{q}_1}{\partial \hat{T}}\)\right|_{\hat{\nu}_1,\hat{\nu}_2}\times (\ref{app3_9})\right]\right. \nn\\
&-&\left.\left . \(\frac{\partial \hat{q}_1}{\partial\hat{\nu}_2}\)\right|_{\hat{T},\hat{\nu}_1}\left[\left . \(\frac{\partial \hat{q}_2}{\partial \hat{T}}\)\right|_{\hat{\nu}_1,\hat{\nu}_2}\times(\ref{app3_7}) - \left . \(\frac{\partial\hat{\e}}{\partial \hat{T}}\)\right|_{\hat{\nu}_1,\hat{\nu}_2}\times (\ref{app3_9})\right]\right\}\,,\label{app3_22}\nn\\
\eea
and to get (\ref{app3_28}),
\be\label{app3_26}
\left. \(\frac{\partial \hat{P}}{\partial\hat{\e}}\)\right|_{\hat{q}_1,\hat{q}_2}\times(\ref{app3_7}) + \left. \(\frac{\partial \hat{P}}{\partial \hat{q}_1}\)\right|_{\hat{\e},\hat{q}_2}\times(\ref{app3_8}) + \left. \(\frac{\partial \hat{P}}{\partial \hat{q}_2}\)\right|_{\hat{\e},\hat{q}_1}\times(\ref{app3_9})\,.
\ee
In order to simplify the RHS and LHS of each of these combinations in turn, we make use of the following, as done equivalently in \cite{Bhattacharya:2011tr}: \emph{Via} the chain rule (for $i=1,2,3$) we get
\bea
\left . \(\frac{\partial \hat{P}}{\partial \Sigma_i}\) \right|_{A_i,B_i} &=& \left . \(\frac{\partial \hat{P}}{\partial\hat{\e}}\) \right|_{\hat{q}_1,\hat{q}_2}\left . \(\frac{\partial\e}{\partial \Sigma_i}\) \right|_{A_i,B_i} + \left . \(\frac{\partial \hat{P}}{\partial \hat{q}_1}\) \right|_{\hat{\e},\hat{q}_2}\left . \(\frac{\partial \hat{q}_1}{\partial \Sigma_i}\) \right|_{A_i,B_i} \nn\\
&&+ \left . \(\frac{\partial \hat{P}}{\partial \hat{q}_2}\) \right|_{\hat{\e},\hat{q}_1}\left . \(\frac{\partial \hat{q}_2}{\partial \Sigma_i}\) \right|_{A_i,B_i}\,, \nn\\
&&\mbox{\ } \label{app3_11}
\eea
where $(\Sigma_1,\Sigma_2,\Sigma_3) = (T,\hat{\nu}_1,\hat{\nu}_2)$, $(A_1,A_2,A_3) = (\hat{\nu}_1, \hat{T},\hat{T})$ and $(B_1,B_2,B_3) = (\hat{\nu}_2,\hat{\nu}_2,\hat{\nu}_1)$, while
\bea\label{app3_10_1}
(\hat{u}\cdot\partial)\Gamma = \left. \(\frac{\partial\Gamma}{\partial \hat{T} }\)\right |_{\hat{\nu}_1,\hat{\nu}_2} (\hat{u}\cdot \partial)\hat{T}+ \left . \(\frac{\partial\Gamma}{\partial \hat{\nu}_1 }\)\right |_{\hat{T},\hat{\nu}_2} (\hat{u}\cdot \partial)\hat{\nu}_1+ \left . \(\frac{\partial\Gamma}{\partial \hat{\nu}_2 }\)\right |_{\hat{T},\hat{\nu}_1}(\hat{u}\cdot \partial)\hat{\nu}_2 \,,\nn\\
&&\mbox{\ }
\eea
where $(\Gamma = \hat{\e}, \hat{q}_1, \hat{q}_2)$. Furthermore, from the first law,
\be\label{app3_13}
\hat{P}+\hat{\e} = \hat{T}(\hat{s}+\hat{q}_1\hat{\nu}_1 + \hat{q}_2 \hat{\nu}_2)\,,\qquad\ud \hat{P} = \(\frac{\hat{P}+\hat{\e}}{\hat{T}}\)\ud \hat{T} + \hat{T} \hat{q}_1\ud \hat{\nu}_1 + \hat{T} \hat{q}_2\ud \hat{\nu}_2\,,
\ee
we know that
\bea
\left.\(\frac{\partial \hat{P}}{\partial \hat{T}}\)\right|_{\hat{\nu}_1,\hat{\nu}_2} = \frac{\hat{P}+\hat{\e}}{\hat{T}}\,,\qquad \left.\(\frac{\partial \hat{P}}{\partial\hat{\nu}_1}\)\right|_{\hat{T},\hat{\nu}_2} = \hat{T} \hat{q}_1\,,\qquad\left.\(\frac{\partial \hat{P}}{\partial\hat{\nu}_2}\)\right|_{\hat{T},\hat{\nu}_1} =\hat{T} \hat{q}_2\,,\label{app3_14}
\eea
and also
\be
\left. \(\frac{\partial\hat{\e}}{\partial\hat{\nu}_1}\) \right|_{\hat{T},\hat{\nu}_2} = \hat{T}^2\left. \(\frac{\partial \hat{q}_1}{\partial \hat{T}}\)\right|_{\hat{\nu}_1,\hat{\nu}_2}\,,\qquad
\left. \(\frac{\partial\hat{\e}}{\partial\hat{\nu}_2}\) \right|_{\hat{T},\hat{\nu}_1} = \hat{T}^2\left. \(\frac{\partial \hat{q}_2}{\partial \hat{T}}\)\right|_{\hat{\nu}_1,\hat{\nu}_2}\,.\label{app3_18}
\ee
We also use the fact that
\be\label{app3_21}
\left. \(\frac{\partial \hat{q}_2}{\partial\hat{\nu}_1}\)\right|_{\hat{T},\hat{\nu}_2} = \left . \(\frac{\partial \hat{q}_1}{\partial \hat{\nu}_2}\)\right |_{\hat{T},\hat{\nu}_1}\,,
\ee
which is a Maxwell relation obtained by using the grand potential density $\vp\equiv \hat{\e} -\hat{T}\hat{s}-\hat{\mu}_1 \hat{q}_1 - \hat{\mu}_2 \hat{q}_2$.

Finally, using (\ref{app3_6}), we may write
\bea
\hat{P}_{ik}\partial_{j}\hat{T}^{jk}
&=& (\hat{P}+\hat{\e})(\hat{u}\cdot \partial)\hat{u}_{i} + \hat{P}_{i}^{\ k}\(\frac{\hat{P}+\hat{\e}}{\hat{T}}\partial_{k}\hat{T} + \hat{T} \hat{q}_1\partial_{k}\hat{\nu}_1 + \hat{T}\hat{q}_2\partial_{k}\hat{\nu}_2\)\,,\nn\\
&&\mbox{\ }\label{app3_29}
\eea
where in the last line above we used (\ref{app3_14}) and the fact that
\be
\partial_{k}\hat{P} = \left.\(\frac{\partial \hat{P}}{\partial \hat{T}}\)\right|_{\hat{\nu}_1,\hat{\nu}_2}\partial_{k}\hat{T} + \left.\(\frac{\partial \hat{P}}{\partial \hat{\nu}_1}\)\right|_{\hat{T},\hat{\nu}_2}\partial_{k}\hat{\nu}_1 + \left.\(\frac{\partial \hat{P}}{\partial \hat{\nu}_2}\)\right|_{\hat{T},\hat{\nu}_1}\partial_{\k}\hat{\nu}_2\,.
\ee
From the equation of motion in (\ref{app3_5}) this is equivalent to
\bea
\hat{P}_{ik}\(F^{(1)jk}_{(0)}\hat{J}_{j}^{(1)} + F^{(2)jk}\hat{J}_{j}^{(2)}\) = \hat{q}_1 F^{(1)j}_{(0)\ \ i}\hat{u}_{j} + \hat{q}_2 F^{(2)j}_{(0)\ \ i}\hat{u}_{j}\,,
\eea
where we used (\ref{app3_6}), so that we get
\be\label{app3_30}
V_{1i}^{(3)}= \frac{\hat{q}_1 \hat{T}}{\hat{P}+\hat{\e}}V_{1i}^{(1)} +  \frac{\hat{q}_2 \hat{T}}{\hat{P}+\hat{\e}}V_{1i}^{(2)}\,.
\ee

With the results in (\ref{app3_20}), (\ref{app3_25}), (\ref{app3_28}) and (\ref{app3_30}) in hand, we may rewrite (\ref{app3_3}) as given in (\ref{ReducedDivEntropyCurrent}). Requiring this to be positive semi-definite allows us to write down the equations from which we may extract the transport coefficients.

\subsection{Computing transport coefficients}\label{app:comptranscoeff}

In this appendix we extract the transport coefficients from (\ref{TensorInv}), and (\ref{BulkViscosityEqn_1}) - (\ref{BulkViscosityEqn}). We first use the conservation equations for the fluid, namely 
\bea
\partial_i \<\hat{T}^{ij}\>_d&=& F^{(1)ij}_{(0)} \<\hat{J}_i^{(1)}\>_d + F^{(2)ij}_{(0)} \<\hat{J}_i^{(2)}\>_d\,,\\
\partial^i \<\hat{J}_i^{(1)}\>_d&=&0\,,\\
\partial^i \<\hat{J}_i^{(2)}\>_d&=&0\,,
\eea
to get 
\bea
\partial_j \log m &=&  \frac{\cosh\omega_1}{\sinh\omega_1}\,\hat u\cdot\partial\w_1~\hat u_j  -\cosh^2\w_1\cosh^2\w_2~ \hat u\cdot\partial\hat u_j \nn\\
&+&\sinh\w_1\cosh\w_1\cosh\w_2~\hat u^iF_{(0)ij}^{(1)} +\cosh^2\w_1\cosh\w_2\sinh\w_2~\hat u^iF_{(0)ij}^{(2)}\,,\nn\\
&&\mbox{\ \ }\nn \\
\hat u\cdot\partial\omega_1&=& \frac{\sinh  \omega_1~\cosh\w_1\cosh^2\w_2}{2(\s-1)\cosh^2\omega_1\cosh^2\w_2+1}\partial\cdot\hat u\,,\nn\\
\hat u\cdot\partial\omega_2&=& \frac{\sinh\w_2\cosh\w_2}{2(\s-1)\cosh^2\omega_1\cosh^2\w_2+1}\partial\cdot\hat u=\frac{\tanh\w_2}{\cosh\w_1\sinh\w_1}\hat u\cdot\partial\omega_1\,,\label{results_0}
\eea
and then we evaluate
\bea
&&\hat{P}^{ij} \<\hat{T}^{\rm diss}_{ij}\>_d = -2\eta_d \left[\cosh\w_1\cosh\w_2\hat{P}^{ij}\partial_{(i}\hat{u}_{j)} -\frac{(d-1)}{(2\s-1)}\(\cosh\w_1\cosh\w_2~\partial\cdot\hat{u} \right. \right. \nn\\
&&\left. \left. \qquad\qquad\qquad+ \cosh\w_2\sinh\w_1~\hat u\cdot\partial\omega_1 + \cosh\w_1\sinh\w_2~ \hat u\cdot\partial\omega_2\)\right] \nn\\
&&\qquad\qquad\ \ \  = -2\eta_d \cosh\w_1\cosh\w_2 \left[1-\frac{(d-1)\cosh^2\w_1\cosh^2\w_2}{2(\s-1)\cosh^2\w_1\cosh^2\w_2+1} \right]\partial\cdot\hat{u}\,,\nn\\
&&\hat{P}_{k}^{i}\hat{P}_{l}^{j} \<\hat{T}^{\rm diss}_{ij}\>_d = -2\eta_d \left[\cosh\w_1\cosh\w_2\hat{P}_{k}^{i}\hat{P}_{l}^{j}\partial_{(i}\hat{u}_{j)} \right. \nn\\
&&\left. \qquad\qquad\qquad\qquad\qquad-\frac{1}{(2\s-1)}\hat{P}_{kl}\(\cosh\w_1\cosh\w_2~\partial\cdot\hat{u} \right. \right. \nn\\
&&\left. \left. \qquad\qquad\qquad\qquad\qquad+ \cosh\w_2\sinh\w_1~\hat u\cdot\partial\omega_1 + \cosh\w_1\sinh\w_2~ \hat u\cdot\partial\omega_2\)\right] \,,\nn \\
&&\hat{P}^{ij} \<\hat{J}_{j}^{(1)\rm diss}\>_d =  -\eta_d\hat{P}^{ij} \left[ \cosh\w_1\partial_j\w_1 -\cosh^3\w_1\cosh\w_2 F_{(0)mj}^{(1)}\hat{u}^{m} \right. \nn\\
&&\left.\qquad\qquad\qquad\qquad\qquad -\cosh^2\w_1\cosh\w_2\sinh\w_1\sinh\w_2 F_{(0)mj}^{(2)}\hat{u}^{m} \right. \nn\\
&&\left. \qquad\qquad\qquad\qquad\qquad+ \cosh^2\w_1\cosh^2\w_2\sinh\w_1\hat{u}\cdot\partial\hat{u}_{j}\right] \,, \nn\\
&&\hat{P}^{ij} \<\hat{J}_{j}^{(2)\rm diss}\>_d =  -\eta_d\hat{P}^{ij} \left[ \sinh\w_1\sinh\w_2\partial_j\w_1+\cosh\w_1\cosh\w_2\partial_j\w_2 \right. \nn\\
&&\left. \qquad\qquad\qquad\qquad\qquad-\cosh^2\w_1\cosh\w_2\sinh\w_1\sinh\w_2 F_{(0)mj}^{(1)}\hat{u}^{m} \right. \nn\\
&&\left. \qquad\qquad\qquad\qquad\qquad -\cosh\w_1\cosh\w_2(1+\cosh^2\w_1\sinh^2\w_2) F_{(0)mj}^{(2)}\hat{u}^{m} \right. \nn\\
&&\left.\qquad\qquad\qquad\qquad\qquad+ \cosh^3\w_1\cosh^2\w_2\sinh\w_2\hat{u}\cdot\partial\hat{u}_{j}\right] \,, \nn\\
&&{\< \cO_{1}^{\rm diss} \>_d}=2\eta_d \left[ \cosh^2\w_1\cosh\w_2\sinh\w_1\sinh^2\w_2\hat u\cdot\partial\omega_1\right. \nn\\
&&\left. \qquad\qquad\qquad\qquad+\cosh^3\w_1\cosh^2\w_2\sinh\w_2\hat u\cdot\partial\omega_2\right. \nn\\
&&\left. \qquad\qquad\qquad-\frac{1}{(2\s-1)}(1+\cosh^2\w_1\sinh^2\w_2)\(\cosh\w_1\cosh\w_2~\partial\cdot\hat{u}\right.\right. \nn\\
 &&\left.\left.\qquad\qquad\qquad\qquad\qquad+ \cosh\w_2\sinh\w_1~\hat u\cdot\partial\omega_1 + \cosh\w_1\sinh\w_2~ \hat u\cdot\partial\omega_2\)\right] \nn\\
&&\qquad\qquad=\< \cO_{3}^{\rm diss} \>_d \,,\nn\\
&&{\< \cO_{2}^{\rm diss} \>_d}=-2\eta_d\frac{1}{(2\s-1)}\(\cosh\w_1\cosh\w_2~\partial\cdot\hat{u} + \cosh\w_2\sinh\w_1~\hat u\cdot\partial\omega_1\right. \nn\\
&&\left. \qquad\qquad\qquad\qquad\qquad\qquad+ \cosh\w_1\sinh\w_2~ \hat u\cdot\partial\omega_2\)\nn\\
&&\qquad\qquad=\< \cO_{3}^{\rm diss} \>_d\,, \nn\\
&&{\< \cO_{3}^{\rm diss} \>_d}=2\eta_d \cosh^2\w_1 \left[\cosh\w_2\sinh\w_1\hat u\cdot\partial\omega_1-\frac{1}{(2\s-1)}\(\cosh\w_1\cosh\w_2~\partial\cdot\hat{u} \right. \right. \nn\\
&&\left. \left. \qquad\qquad\qquad+ \cosh\w_2\sinh\w_1~\hat u\cdot\partial\omega_1 + \cosh\w_1\sinh\w_2~ \hat u\cdot\partial\omega_2\)\right]\nn\\
&&\qquad\qquad=-2\eta_d \frac{\cosh^3\w_1\cosh^3\w_2}{2(\s-1)\cosh^2\w_1\cosh^2\w_2+1}\partial\cdot\hat{u}\,, \label{Op1}\\
&&{\< \cO_{4}^{\rm diss} \>_d}=2\eta_d \left[ \frac{1}{2}\cosh\w_1\cosh\w_2\sinh\w_2\cosh(2\w_1)\hat u\cdot\partial\omega_1 \right. \nn\\
&&\left.\qquad\qquad\qquad\qquad+\frac{1}{2}\cosh^2\w_1\cosh^2\w_2\sinh\w_1\hat u\cdot\partial\omega_2 \right. \nn\\
&&\left. \qquad\qquad\qquad\qquad-\frac{1}{(2\s-1)}\cosh\w_1\sinh\w_1\sinh\w_2\(\cosh\w_1\cosh\w_2~\partial\cdot\hat{u} \right.\right. \nn\\
&&\left.\left. \qquad\qquad\qquad\qquad\qquad+ \cosh\w_2\sinh\w_1~\hat u\cdot\partial\omega_1 + \cosh\w_1\sinh\w_2~ \hat u\cdot\partial\omega_2\)\right] \nn\\
&&\qquad\qquad =0 \,, \nn\\
&&{\< \cO_{\y}^{\rm diss} \>_d}={\< \cO_{3}^{\rm diss} \>_d} \,, \nn\\
&&{\< \cO_{\z}^{\rm diss} \>_d}={\< \cO_{\x}^{\rm diss} \>_d}={\< \cO_{A_{(0)}^{(1)}}^{\rm diss} \>_d}=0\,.\label{results}
\eea

Evaluating $\hat{\z}_s$ from (\ref{BulkViscosityEqn}) also requires us to calculate the variations of $\hat{P}$ with respect to $\hat{\e}$, $\hat{q}_1$ and $\hat{q}_2$ respectively, for which we use 
\bea
\ud \hat q_1=0\ \mbox{and}\ \ud \hat q_2=0 &\Rightarrow& \ud\omega_1=-\frac{2\s\sinh\w_1\cosh\w_1\cosh^2\w_2}{(2\cosh^2\w_1\cosh^2\w_2 -1)}\frac{\ud m}m\,, \nn\\
&& \ud\omega_2=-\frac{2\s\cosh\w_2\sinh\w_2}{(2\cosh^2\w_1\cosh^2\w_2 -1)}\frac{\ud m}m\,, \nn\\
\ud \hat \e=0\ \mbox{and}\ \ud \hat q_2=0 &\Rightarrow& \ud\omega_1=-\frac{(2\s\cosh^2\w_1\cosh^2\w_2-2\cosh^2\w_2+1)}{2\cosh\w_1\sinh\w_1\cosh^2\w_2}\frac{\ud m}m \,,\nn\\
&& \ud\omega_2=-\frac{\sinh\w_2}{\cosh^2\w_1\cosh\w_2}\frac{\ud m}m\,, \nn\\
\ud \hat \e=0\ \mbox{and}\ \ud \hat q_1=0 &\Rightarrow& \ud\omega_1=-\frac{\sinh\w_1(2\s\cosh^2\w_1\cosh^2\w_2+1)}{2\cosh^3\w_1\cosh^2\w_2}\frac{\ud m}m\,, \nn\\
&& \ud\omega_2=-\frac{(2\s\cosh^2\w_1\cosh^2\w_2-2\cosh^2\w_1+1)}{2\cosh^4\w_1\cosh\w_2\sinh\w_2}\frac{\ud m}m\,. \nn
\eea
From these relations it follows that
\bea
\frac{\partial\hat{P}}{\partial\hat{\e}}&\equiv&\left.\frac{\partial\hat{P}}{\partial\hat{\e}}\right|_{\hat{q}_1,\hat{q}_2}=\frac{(2\cosh^2\w_1\cosh^2\w_2 -1)}{(2(\s-1)\cosh^2\w_1\cosh^2\w_2 +1)}\,, \nn\\
\frac{\partial\hat{P}}{\partial\hat{q}_1}&\equiv&\left.\frac{\partial\hat{P}}{\partial\hat{q}_1}\right|_{\hat{\e},\hat{q}_2}=-\frac{2\cosh\w_1\sinh\w_1\cosh\w_2}{(2(\s-1)\cosh^2\w_1\cosh^2\w_2 +1)} \,,\nn\\
\frac{\partial\hat{P}}{\partial\hat{q}_2}&\equiv&\left.\frac{\partial\hat{P}}{\partial\hat{q}_2}\right|_{\hat{\e},\hat{q}_1}=-\frac{2\cosh^2\w_1\cosh\w_2\sinh\w_2}{(2(\s-1)\cosh^2\w_1\cosh^2\w_2 +1)} \,.  \label{Pderivs}
\eea
Using $\< \cO_{1}^{\rm diss} \>_d=\< \cO_{3}^{\rm diss} \>_d$ and $\< \cO_{4}^{\rm diss} \>_d=0$, the equation from which we may extract the bulk viscosity (\ref{BulkViscosityEqn}) becomes
\bea
&&\frac{\hat{P}^{ij} \<\hat{T}^{\rm diss}_{ij}\>_d}{(d -1)}+\< \cO_{3}^{\rm diss} \>_d\left[ \frac{\partial\hat{P}}{\partial\hat{\e}}\left( \frac{\tanh^2\w_1}{\cosh^2\w_2}+\tanh^2\w_2\right) \right. \nn\\
&&\left. \qquad\qquad\qquad\qquad\qquad+ \frac{\partial\hat{P}}{\partial\hat{q}_1}\frac{\tanh\w_1}{\cosh\w_2} + \frac{\partial\hat{P}}{\partial\hat{q}_2}\tanh\w_2 \right] = -\hat{\z}_s ~\partial\cdot\hat{u} \,.  \label{bulkvisceqn}
\eea
Plugging (\ref{Op1}) and (\ref{Pderivs}) into the equation above yields the bulk viscosity (\ref{BulkViscosity}), while substituting the results in (\ref{results_0}) and (\ref{results}) into the other transport coefficient equations, namely (\ref{TensorInv}), (\ref{BulkViscosityEqn_1}) and (\ref{BulkViscosityEqn_2}), gives (\ref{ShearViscosity}) - (\ref{HeatConductivity_3}).

\subsection{Derivation of formula for bulk to shear viscosity ratio}\label{app:derivratio}

Equation (\ref{ElingOzFormula2}) was derived in \cite{Eling:2011ms}, and expresses the bulk viscosity to shear viscosity ratio in terms of the dependence of scalar fields at the horizon on thermodynamic variables such as entropy and charge densities. It is derived using the null focusing (Raychaudhuri) equation, which is equivalent \emph{via} the fluid/gravity correspondence to the entropy/balance law of the fluid.

In \cite{Eling:2011ms} the action under consideration is a $(d+1)$-dimensional gravitational action in the Einstein frame, where the kinetic terms for the various scalar fields are canonically normalized. However, we are interested in action (\ref{finact}), where the axion kinetic term is clearly not canonically normalized, so we cannot simply apply (\ref{ElingOzFormula2}) to our case. In this appendix we thus repeat the derivation given in \cite{Eling:2011ms} (while also adjusting the notation to match our conventions) for a general action of the form 
\bea\label{genact}
S_{(d+1)}&=& L \int \ud ^{d+1} x \sqrt{-\bar{g}_{(d+1)}} \left[ \bar{R} -\frac{1}{2}\Omega_{IJ}(\phi)\partial_{M}\phi^{I}\partial^{M}\phi^{J} -V(\phi^{I})\right] + S_{gauge}\,, \nn\\
&&\mbox{\ } 
\eea
where $\Omega_{IJ}$ parametrizes the normalization of the scalar field kinetic terms. In \cite{Eling:2011ms}, $\Omega_{IJ}=\mbox{diag}\{1,1,\cdots,1\}$, while in our case $\phi^{I}=\{\bar{\y},\bar{\z},\bar{\x},A_{(0)}^{(1)}\}$ and 
\be
\Omega_{IJ}=\mbox{diag}\{1,1,1,\mbox{exp}(c_1\bar{\x}-c_2\bar{\z})\}\,, 
\ee
where
\be
c_1 = \sqrt{\frac{2(2\s-d-2)}{(2\s-d-1)}}\,,\qquad c_2=\sqrt{\frac{2(2\s-d)}{(2\s-d-1)}}\,.
\ee
The derivation in \cite{Eling:2011ms} follows through exactly, except that all occurrences of $\sum_{I}(\partial\phi^{I})^{2}$ (or equivalent expressions) are replaced by $\Omega_{IJ}\phi^{I}\phi^{J}$, in particular the Raychaudhuri equation. 

The procedure involves considering the focusing equation at subsequent orders in derivatives of $z^{i}$, the local coordinates on the horizon (with $x^{M}=(\rho,z^{i})$, $M=0,\cdots,d$, and $\r$ a transverse coordinate, with $\rho=\rho_h$ on the horizon). Following the procedure outlined in \cite{Eling:2011ms}, we are ultimately lead to the Raychaudhuri equation to second order
\bea\label{Raysecord}
\partial_{i}(\hat{s}\hat{u}^{i}) = \frac{\hat{s}}{2\pi \hat{T}}\hat{\s}_{ij}\hat{\s}^{ij} +\frac{\hat{s}}{4\pi \hat{T}}\Omega_{IJ}^{h}\(\hat{s}\frac{\ud\phi_{h}^{I}}{\ud\hat{s}} + \hat{q}^a\frac{\ud\phi_{h}^{I}}{\ud\hat{q}^a}\)\(\hat{s}\frac{\ud\phi_{h}^{J}}{\ud\hat{s}} + \hat{q}^a\frac{\ud\phi_{h}^{J}}{\ud\hat{q}^a}\)(\partial_{p}\hat{u}^{p})^{2}\,, \nn\\
&&\mbox{\ \ }
\eea
where $\hat{s}$ is the entropy density, $\hat{T}$ is the temperature, $\hat{q}^a$ are charge densities, all quantities with subscript (or superscript) $h$ are evaluated at the horizon, $\hat{u}^{M}=(0,\hat{u}^{i})$ is a vector field obtained by raising the null cotangent vector to the horizon with the bulk metric, and 
\be
\hat{\s}_{ij}=\hat{P}_{i}^{m}\hat{P}_{j}^{n}\partial_{(m}\hat{u}_{n)}-\frac{1}{(d-1)}\hat{P}_{ij}\partial_{p}\hat{u}^{p}\,, \qquad\hat{P}_{ij}=\hat{\eta}_{ij}+\hat{u}_{i}\hat{u}_{j}\,. \nn
\ee
(\ref{Raysecord}) has the form of a fluid entropy balance law
\be
\partial_{i}(\hat{s}\hat{u}^{i}) =\frac{2\hat{\eta}}{\hat{T}}\hat{\s}_{ij}\hat{\s}^{ij}+\frac{\hat{\z}_s}{\hat{T}}(\partial_{p}\hat{u}^{p})^{2}\,,
\ee
\emph{via} the fluid/gravity correspondence (the shear viscosity obeys $\hat{\eta}/\hat{s}=1/4\pi$). We may thus extract the bulk to shear viscosity ratio, which is 
\be
\frac{\hat{\z}_s}{\hat{\eta}}=\Omega_{IJ}^{h}\(\hat{s}\frac{\ud\phi_{h}^{I}}{\ud\hat{s}} + \hat{q}^a\frac{\ud\phi_{h}^{I}}{\ud\hat{q}^a}\)\(\hat{s}\frac{\ud\phi_{h}^{J}}{\ud\hat{s}} + \hat{q}^a\frac{\ud\phi_{h}^{J}}{\ud\hat{q}^a}\)\,.
\ee
In our case, $\Omega_{IJ}^{h}=\left.\mbox{diag}\{1,1,1,\mbox{exp}(c_1\bar{\x}-c_2\bar{\z})\}\right|_{h}$, and the horizon occurs at $\rho=m^{-2}$. Now,
\be
\left.e^{c_1\bar{\x}-c_2\bar{\z}}\right|_{h} = e^{c_1\bar{\x}_{h}-c_2\bar{\z}_{h}} = \(\frac{\cosh\w_1}{\cosh\w_2}\)^{2}\,,
\ee
where
\bea\label{horzx}
\bar{\x}_{h} &=& \sqrt{\frac{2(2\s-d-2)}{(2\s-d-1)}}\ln\cosh\w_1\,, \nn\\ 
\bar{\z}_{h} &=& \sqrt{\frac{2}{(2\s-d)(2\s-d-1)}}\ln \(\frac{(\cosh\w_{2})^{2\s-d-1}}{\cosh\w_1}\)\,.
\eea
In (\ref{horzx}) above we used (\ref{NormScalars}) and (\ref{ScalarFields}) evaluated at $\rho = m^{-2}$.

\subsection{Checking bulk to shear viscosity ratio}\label{app:evalratio}

Having derived the formula (\ref{ElingOzFormula1}) in appendix \ref{app:derivratio}, we now re-evaluate the bulk to shear viscosity ratio given in (\ref{ratio_1}) using this formula.

The charge densities and entropy are given by (\ref{thermo}) and (\ref{shat}), and the scalars are
\bea
\bar{\psi}_h \equiv \bar{\psi}(\r = m^{-2}) &=& \sqrt{\frac{2(2\s-1)}{(2\s-d)(d-1)}}\left.\log\left[ \frac{1}{\r^{\s-d/2}}K_1(\r)^{1/2}K_2(\r)^{1/2} \right]\right|_{\r=m^{-2}}\nn\\
&=& \sqrt{\frac{2(2\s-1)}{(2\s-d)(d-1)}}\log[m^{2\s-d}\cosh\w_1\cosh\w_2]\,, \nn\\
\bar{\z}_h\equiv \bar{\z}(\r = m^{-2}) &=& \sqrt{\frac{2}{(2\s-d)(2\s-d-1)}}\left.\log\left[ K_2(\r)^{\frac{1}{2}(2\s-d-1)}K_1(\r)^{-1/2} \right]\right |_{\r=m^{-2}} \nn\\
&=&\sqrt{\frac{2}{(2\s-d)(2\s-d-1)}}(2\s-d-2)\log[\cosh\w_1]\,, \nn\\
\bar{\x}_h\equiv\bar{\xi}(\r = m^{-2}) &=& \sqrt{\frac{2}{(2\s-d-1)(2\s-d-2)}}\left.\log \left[ K_1(\r)^{\frac{1}{2}(2\s-d-2)}\right]\right |_{\r=m^{-2}} \nn\\
&=&\sqrt{\frac{2}{(2\s-d-1)(2\s-d-2)}}\log\left[\frac{\cosh\w_2^{(2\s-d-1)}}{\cosh\w_1}\right]\,,
\eea
as obtained from (\ref{ScalarFields}) and (\ref{bbraneconfdt2}) with canonical normalization (\ref{NormScalars}), and further evaluated at the horizon. The axion is given by (\ref{axion})
\be
A^{(0)}_h \equiv \left.((K_1^{\prime}(\r))^{-1}-1)\sinh\w_2\right |_{\r=m^{-2}} = -\frac{\sinh\w_1\sinh\w_2}{\cosh\w_1}\,.
\ee
Notice that
\bea
\ud \hat q_1=0\ \mbox{and}\ \ud \hat q_2=0 &\Rightarrow& \ud\omega_1=-\frac{2\s\sinh\w_1\cosh\w_1\cosh^2\w_2}{(2\cosh^2\w_1\cosh^2\w_2 -1)}\frac{\ud m}m\,, \nn\\
&& \ud\omega_2=-\frac{2\s\cosh\w_2\sinh\w_2}{(2\cosh^2\w_1\cosh^2\w_2 -1)}\frac{\ud m}m\,, \nn\\
\ud \hat s=0\ \mbox{and}\ \ud \hat q_2=0 &\Rightarrow& \ud\omega_1= -\frac{\cosh\w_1}{\sinh\w_1}(2(\s-1)\cosh^2\w_2+1)\frac{\ud m}m\,, \nn\\
&& \ud\omega_2=2(\s-1)\cosh\w_2\sinh\w_2\frac{\ud m}m\,, \nn\\
\ud \hat s=0\ \mbox{and}\ \ud \hat q_1=0 &\Rightarrow& \ud\omega_1=-\frac{\sinh\w_1}{\cosh\w_1}\frac{\ud m}m\,, \nn\\
&& \ud\omega_2=-\frac{\cosh\w_2(2(\s-1)\cosh^2\w_1+1)}{\cosh^2\w_1\sinh\w_2}\frac{\ud m}m\,, \nn
\eea
from which it is then straightforward  to derive 
\bea
\ud(\log\hat  s)|_{\hat q_1,\hat{q}_2} &=& \left .\frac{\ud \hat{s}}{\hat{s}}\right |_{\hat q_1,\hat{q}_2} = \frac{(2(\s-1)\cosh^2\omega_1\cosh^2\w_2 +1)}{(2\cosh^2\w_1\cosh^2\w_2 -1)}\frac{\ud m}m\,,\nn\\
\ud(\log\hat q_1)|_{\hat s,\hat{q}_2} &=& \left. \frac{\ud \hat{q}_1}{\hat{q}_1}\right|_{\hat s,\hat{q}_2} = -\frac{(2(\s-1)\cosh^2\omega_1\cosh^2\w_2 +1)}{\sinh^2\w_1} \frac{\ud m}m\,,\nn \\
\ud(\log\hat q_2)|_{\hat s,\hat{q}_1} &=&\left . \frac{\ud \hat{q}_2}{\hat{q}_2}\right|_{\hat s,\hat{q}_1}= -\frac{(2(\s-1)\cosh^2\omega_1\cosh^2\w_2 +1)}{\cosh^2\w_1\sinh^2\w_2} \frac{\ud m}m\,,\nn \\
\ud( \bar{\y}_h)|_{\hat q_1,\hat q_2} &=&\sqrt{\frac{2(2\s-1)}{(d-1)(2\s-d)}}\frac{(2(\s-d)\cosh^2\omega_1\cosh^2\w_2~+d)}{(2\cosh^2\w_1\cosh^2\w_2-1)} \frac{\ud m}m\,,\nn \\
\ud( \bar{\y}_h)|_{\hat s,\hat q_2} &=&-\sqrt{\frac{2(2\s-1)(d-1)}{(2\s-d)}}\frac{\ud m}m\,, \nn\\
\ud( \bar{\y}_h)|_{\hat s,\hat q_1} &=&-\sqrt{\frac{2(2\s-1)(d-1)}{(2\s-d)}}\frac{\ud m}m\,, \nn\\
\ud( \bar{\z}_h)|_{\hat q_1,\hat q_2} &=&\sqrt{\frac{2}{(2\s-d)(2\s-d-1)}}\nn\\
&&\times\frac{2\s(\cosh^2\w_1\cosh^2\w_2-(2\s-d)\sinh^2\w_2-1)}{(2\cosh^2\w_1\cosh^2\w_2-1)}\frac{\ud m}m\,, \nn\\
\ud( \bar{\z}_h)|_{\hat s,\hat q_2} &=&\sqrt{\frac{2}{(2\s-d)(2\s-d-1)}}(2(\s-1)(2\s-d)\sinh^2\w_2+2\s-1) \frac{\ud m}m\,, \nn\\
\ud( \bar{\z}_h)|_{\hat s,\hat q_1} &=&\sqrt{\frac{2}{(2\s-d)(2\s-d-1)}}\nn\\
&&\times\frac{(-(2\s-d)(2(\s-1)\cosh^2\w_1+1)+(2\s-1)\cosh^2\w_1)}{\cosh^2\w_1}\frac{\ud m}m\,, \nn\\
\ud( \bar{\x}_h)|_{\hat q_1,\hat q_2} &=&-\sqrt{\frac{2(2\s-d-2)}{(2\s-d-1)}}\frac{2\s\sinh^2\w_1\cosh^2\w_2}{(2\cosh^2\w_1\cosh^2\w_2-1)}\frac{\ud m}m\,, \nn\\
\ud( \bar{\x}_h)|_{\hat s,\hat q_2} &=&-\sqrt{\frac{2(2\s-d-2)}{(2\s-d-1)}}(2(\s-1)\cosh^2\w_2+1)\frac{\ud m}m\,, \nn\\
\ud( \bar{\x}_h)|_{\hat s,\hat q_1} &=&-\sqrt{\frac{2(2\s-d-2)}{(2\s-d-1)}}\frac{\sinh^2\w_1}{\cosh^2\w_1}\frac{\ud m}m\,, \nn\\
\ud( A_{h}^{(0)})|_{\hat q_1,\hat q_2} &=&\frac{4\s \sinh\w_1\sinh\w_2\cosh^2\w_2}{\cosh\w_1(2\cosh^2\w_1\cosh^2\w_2-1)}\frac{\ud m}m\,,\nn\\
\ud( A_{h}^{(0)})|_{\hat s,\hat q_2} &=&-\frac{\sinh\w_2}{\sinh\w_1\cosh\w_1}\(2(\s-1)\cosh^2\w_1\cosh^2\w_2\right. \nn\\
&&\left.\qquad\qquad\qquad\qquad\qquad-4(\s-1)\cosh^2\w_2 -1\)\frac{\ud m}m\,,\nn\\
\ud( A_{h}^{(0)})|_{\hat s,\hat q_1} &=&\frac{\sinh\w_1}{\sinh\w_2\cosh^3\w_1}\(2(\s-1)\cosh^2\w_1\cosh^2\w_2 \right. \nn\\
&&\left. \qquad\qquad\qquad\qquad\qquad+2\cosh^2\w_2 -1\)\frac{\ud m}m\,. \label{Diffs}\nn\\
\eea
We substitute the values in (\ref{Diffs}) into
\bea
\frac{\hat{\z}_s}{\hat{\eta}} &=&\left(\left.\frac{\ud \bar{\y}_h}{\ud \log\hat{s}}\right|_{\hat{q}_1,\hat{q}_2}+\left.\frac{\ud \bar{\y}_h}{\ud \log\hat{q}_1}\right|_{\hat{s},\hat{q}_2}+\left.\frac{\ud \bar{\y}_h}{\ud \log\hat{q}_2}\right|_{\hat{s},\hat{q}_1}\right)^2 \nn\\
&+&\left(\left.\frac{\ud \bar{\z}_h}{\ud \log\hat{s}}\right|_{\hat{q}_1,\hat{q}_2}+\left.\frac{\ud \bar{\z}_h}{\ud \log\hat{q}_1}\right|_{\hat{s},\hat{q}_2}+\left.\frac{\ud \bar{\z}_h}{\ud \log\hat{q}_2}\right|_{\hat{s},\hat{q}_1}\right)^2 \nn\\
&+&\left(\left.\frac{\ud \bar{\x}_h}{\ud \log\hat{s}}\right|_{\hat{q}_1,\hat{q}_2}+\left.\frac{\ud \bar{\x}_h}{\ud \log\hat{q}_1}\right|_{\hat{s},\hat{q}_2}+\left.\frac{\ud \bar{\x}_h}{\ud \log\hat{q}_2}\right|_{\hat{s},\hat{q}_1}\right)^2 \nn\\
&+&\(\frac{\cosh\w_1}{\cosh\w_2}\)^2\left(\left.\frac{\ud A^{(0)}_{h}}{\ud \log\hat{s}}\right|_{\hat{q}_1,\hat{q}_2}+\left.\frac{\ud A^{(0)}_{h}}{\ud \log\hat{q}_1}\right|_{\hat{s},\hat{q}_2}+\left.\frac{\ud A^{(0)}_{h}}{\ud \log\hat{q}_2}\right|_{\hat{s},\hat{q}_1}\right)^2\,, \nn
\eea
and obtain precisely the ratio as evaluated from (\ref{ShearViscosity}) and (\ref{BulkViscosity}).

\bibliographystyle{JHEP}

\end{document}